\documentclass{ws-mpla}

\usepackage{tikz}
\definecolor{maroon}{RGB}{128,0,0}
\definecolor{navy}{RGB}{0,0,128}
\definecolor{midnightblue}{RGB}{0,51,102}
\definecolor{darkblue}{RGB}{0,0,139}

\newcommand{\href}[2]{#2}
\usepackage{relsize}
\usepackage[caption=false,labelformat=simple]{subfig}

\usetikzlibrary{calc, positioning, snakes}
\tikzset{
  baseline={([yshift=-0.75ex]current bounding box.center)},
  zerosep/.style={inner sep=0pt, outer sep=0pt, minimum size=0pt},
  node distance=8pt,
  align at top/.style={baseline=(current bounding box.north)},
}

\usetikzlibrary{decorations,decorations.markings}
\tikzset{
  strike out/.style={
    postaction=decorate,
    decoration={
      markings,
      mark=at position 0.5 with {
        \draw[-] (-2pt, -3pt) -- (2pt, 3pt);
      }
    }
  }
}

\tikzset{
vev/.style={strike out},
pdvev/.style={snake=snake},
}

\newcommand{\tf}{\mathfrak{t}}

\newcommand{\del}{\partial}

\newcommand{\id}{\mathop{\mathrm{id}}\nolimits}

\newcommand{\ket}[1]{|#1\rangle}

\newcommand{\biggvev}[1]{\biggl\langle #1 \biggr\rangle}

\newcommand{\diag}{\mathop{\mathrm{diag}}\nolimits}

\newcommand{\Tr}{\mathop{\mathrm{Tr}}\nolimits}

\newcommand{\SU}{\mathrm{SU}}

\newcommand{\slf}{\mathfrak{sl}}

\newcommand{\U}{\mathrm{U}}

\newcommand{\R}{\mathbb{R}}

\let\nc\newcommand
\let\renc\renewcommand
\nc{\wbar}{\overline}
\let\td\tilde
\let\wtd\widetilde
\let\wht\widehat
\let\mcl\mathcal

\nc{\ab}{{\bar{a}}} \nc{\at}{\tilde{a}} \nc{\ah}{\hat{a}}
\nc{\bb}{{\bar{b}}} \nc{\bt}{\tilde{b}} \nc{\bh}{\hat{b}}
\nc{\cb}{{\bar{c}}} \nc{\ct}{\tilde{c}} 
\nc{\db}{{\bar{d}}} \nc{\dt}{\tilde{d}} \renc{\dh}{\hat{d}}
\nc{\eb}{{\bar{e}}} \nc{\et}{\tilde{e}} \nc{\eh}{\hat{e}}
\nc{\fb}{{\bar{f}}} \nc{\ft}{\tilde{f}} \nc{\fh}{\hat{f}}
\nc{\gb}{{\bar{g}}} \nc{\gt}{\tilde{g}} \nc{\gh}{\hat{g}}
\nc{\hb}{{\bar{h}}} \nc{\hh}{\hat{h}} 
\nc{\ib}{{\bar{\imath}}} \nc{\ih}{\hat{\imath}} 
\nc{\jb}{{\bar{\jmath}}} \nc{\jt}{\tilde{\jmath}} \nc{\jh}{\hat{\jmath}}
\nc{\kb}{{\bar{k}}} \nc{\kt}{\tilde{k}} \nc{\kh}{\hat{k}}
\nc{\lb}{{\bar{l}}} \nc{\lt}{\tilde{l}} \nc{\lh}{\hat{l}}
\nc{\mb}{{\bar{m}}} \nc{\mt}{\tilde{m}} \nc{\mh}{\hat{m}}
\nc{\nb}{{\bar{n}}} \nc{\nt}{\tilde{n}} \nc{\nh}{\hat{n}}
\nc{\ob}{{\bar{o}}} \nc{\ot}{\tilde{o}} \nc{\oh}{\hat{o}}
\nc{\pb}{{\bar{p}}} \nc{\pt}{\tilde{p}} \nc{\ph}{\hat{p}}
\nc{\qb}{{\bar{q}}} \nc{\qt}{\tilde{q}} \nc{\qh}{\hat{q}}
\nc{\rb}{{\bar{r}}} \nc{\rt}{\tilde{r}} \nc{\rh}{\hat{r}}
\renc{\sb}{{\bar{s}}} \nc{\st}{\tilde{s}} \nc{\sh}{\hat{s}}
\nc{\tb}{{\bar{t}}} \renc{\th}{\hat{t}} 
\nc{\ub}{{\bar{u}}} \nc{\ut}{\tilde{u}} \nc{\uh}{\hat{u}}
\nc{\vb}{{\bar{v}}} \nc{\vt}{\tilde{v}} \nc{\vh}{\hat{v}}
\nc{\wb}{{\bar{w}}} \nc{\wt}{\tilde{w}} \nc{\wh}{\hat{w}}
\nc{\xb}{{\bar{x}}} \nc{\xt}{\tilde{x}} \nc{\xh}{\hat{x}}
\nc{\yb}{{\bar{y}}} \nc{\yt}{\tilde{y}} \nc{\yh}{\hat{y}}
\nc{\zb}{{\bar{z}}} \nc{\zt}{\tilde{z}} \nc{\zh}{\hat{z}}

\nc{\Ab}{{\wbar{A}}} \nc{\At}{{\wtd{A}}} \nc{\Ah}{{\wht{A}}}
\nc{\Bb}{{\wbar{B}}} \nc{\Bt}{{\wtd{B}}} \nc{\Bh}{{\wht{B}}}
\nc{\Cb}{{\wbar{C}}} \nc{\Ct}{{\wtd{C}}} \nc{\Ch}{{\wht{C}}}
\nc{\Db}{{\wbar{D}}} \nc{\Dt}{{\wtd{D}}} \nc{\Dh}{{\wht{D}}}
\nc{\Eb}{{\wbar{E}}} \nc{\Et}{{\wtd{E}}} \nc{\Eh}{{\wht{E}}}
\nc{\Fb}{{\wbar{F}}} \nc{\Ft}{{\wtd{F}}} \nc{\Fh}{{\wht{F}}}
\nc{\Gb}{{\wbar{G}}} \nc{\Gt}{{\wtd{G}}} \nc{\Gh}{{\wht{G}}}
\nc{\Hb}{{\wbar{H}}} \nc{\Ht}{{\wtd{H}}} \nc{\Hh}{{\wht{H}}}
\nc{\Ib}{{\bar{I}}} \nc{\It}{{\wtd{I}}} \nc{\Ih}{{\wht{I}}}
\nc{\Jb}{{\bar{J}}} \nc{\Jt}{{\wtd{J}}} \nc{\Jh}{{\wht{J}}}
\nc{\Kb}{{\wbar{K}}} \nc{\Kt}{{\wtd{K}}} \nc{\Kh}{{\wht{K}}}
\nc{\Lb}{{\wbar{L}}} \nc{\Lt}{{\wtd{L}}} \nc{\Lh}{{\wht{L}}}
\nc{\Mb}{{\wbar{M}}} \nc{\Mt}{{\wtd{M}}} \nc{\Mh}{{\wht{M}}}
\nc{\Nb}{{\wbar{N}}} \nc{\Nt}{{\wtd{N}}} \nc{\Nh}{{\wht{N}}}
\nc{\Ob}{{\wbar{O}}} \nc{\Ot}{{\wtd{O}}} \nc{\Oh}{{\wht{O}}}
\nc{\Pb}{{\wbar{P}}} \nc{\Pt}{{\wtd{P}}} \nc{\Ph}{{\wht{P}}}
\nc{\Qb}{{\wbar{Q}}} \nc{\Qt}{{\wtd{Q}}} \nc{\Qh}{{\wht{Q}}}
\nc{\Rb}{{\wbar{R}}} \nc{\Rt}{{\wtd{R}}} \nc{\Rh}{{\wht{R}}}
\nc{\Sb}{{\wbar{S}}} \nc{\St}{{\wtd{S}}} \nc{\Sh}{{\wht{S}}}
\nc{\Tb}{{\wbar{T}}} \nc{\Tt}{{\wtd{T}}} \nc{\Th}{{\wht{T}}}
\nc{\Ub}{{\wbar{U}}} \nc{\Ut}{{\wtd{U}}} \nc{\Uh}{{\wht{U}}}
\nc{\Vb}{{\wbar{V}}} \nc{\Vt}{{\wtd{V}}} \nc{\Vh}{{\wht{V}}}
\nc{\Wb}{{\wbar{W}}} \nc{\Wt}{{\wtd{W}}} \nc{\Wh}{{\wht{W}}}
\nc{\Xb}{{\wbar{X}}} \nc{\Xt}{{\wtd{X}}} \nc{\Xh}{{\wht{X}}}
\nc{\Yb}{{\wbar{Y}}} \nc{\Yt}{{\wtd{Y}}} \nc{\Yh}{{\wht{Y}}}
\nc{\Zb}{{\wbar{Z}}} \nc{\Zt}{{\wtd{Z}}} \nc{\Zh}{{\wht{Z}}}

\nc{\CA}{{\mcl{A}}} \nc{\CAb}{{\wbar{\CA}}} \nc{\CAt}{{\wtd{\CA}}} \nc{\CAh}{{\wht{\CA}}}
\nc{\CB}{{\mcl{B}}} \nc{\CBb}{{\wbar{\CB}}} \nc{\CBt}{{\wtd{\CB}}} \nc{\CBh}{{\wht{\CB}}}
\nc{\CC}{{\mcl{C}}} \nc{\CCb}{{\wbar{\CC}}} \nc{\CCt}{{\wtd{\CC}}} \nc{\CCh}{{\wht{\CC}}}
\nc{\cD}{{\mcl{D}}} \nc{\cDb}{{\wbar{\cD}}} \nc{\cDt}{{\wtd{\cC}}} \nc{\cDh}{{\wht{\cD}}}
\nc{\CE}{{\mcl{E}}} \nc{\CEb}{{\wbar{\CE}}} \nc{\CEt}{{\wtd{\CE}}} \nc{\CEh}{{\wht{\CE}}}
\nc{\CF}{{\mcl{F}}} \nc{\CFb}{{\wbar{\CF}}} \nc{\CFt}{{\wtd{\CF}}} \nc{\CFh}{{\wht{\CF}}}
\nc{\CG}{{\mcl{G}}} \nc{\CGb}{{\wbar{\CG}}} \nc{\CGt}{{\wtd{\CG}}} \nc{\CGh}{{\wht{\CG}}}
\nc{\CH}{{\mcl{H}}} \nc{\CHb}{{\wbar{\CH}}} \nc{\CHt}{{\wtd{\CH}}} \nc{\CHh}{{\wht{\CH}}}
\nc{\CI}{{\mcl{I}}} \nc{\CIb}{{\wbar{\CI}}} \nc{\CIt}{{\wtd{\CI}}} \nc{\CIh}{{\wht{\CI}}}
\nc{\CJ}{{\mcl{J}}} \nc{\CJb}{{\wbar{\CJ}}} \nc{\CJt}{{\wtd{\CJ}}} \nc{\CJh}{{\wht{\CJ}}}
\nc{\CK}{{\mcl{K}}} \nc{\CKb}{{\wbar{\CK}}} \nc{\CKt}{{\wtd{\CK}}} \nc{\CKh}{{\wht{\CK}}}
\nc{\CL}{{\mcl{L}}} \nc{\CLb}{{\wbar{\CL}}} \nc{\CLt}{{\wtd{\CL}}} \nc{\CLh}{{\wht{\CL}}}
\nc{\CM}{{\mcl{M}}} \nc{\CMb}{{\wbar{\CM}}} \nc{\CMt}{{\wtd{\CM}}} \nc{\CMh}{{\wht{\CM}}}
\nc{\CN}{{\mcl{N}}} \nc{\CNb}{{\wbar{\CN}}} \nc{\CNt}{{\wtd{\CN}}} \nc{\CNh}{{\wht{\CN}}}
\nc{\CO}{{\mcl{O}}} \nc{\COb}{{\wbar{\CO}}} \nc{\COt}{{\wtd{\CO}}} \nc{\COh}{{\wht{\CO}}}
\nc{\CP}{{\mcl{P}}} \nc{\CPb}{{\wbar{\CP}}} \nc{\CPt}{{\wtd{\CP}}} \nc{\CPh}{{\wht{\CP}}}
\nc{\CQ}{{\mcl{Q}}} \nc{\CQb}{{\wbar{\CQ}}} \nc{\CQt}{{\wtd{\CQ}}} \nc{\CQh}{{\wht{\CQ}}}
\nc{\CR}{{\mcl{R}}} \nc{\CRb}{{\wbar{\CR}}} \nc{\CRt}{{\wtd{\CR}}} \nc{\CRh}{{\wht{\CR}}}
\nc{\CS}{{\mcl{S}}} \nc{\CSb}{{\wbar{\CS}}} \nc{\CSt}{{\wtd{\CS}}} \nc{\CSh}{{\wht{\CS}}}
\nc{\CT}{{\mcl{T}}} \nc{\CTb}{{\wbar{\CT}}} \nc{\CTt}{{\wtd{\CT}}} \nc{\CTh}{{\wht{\CT}}}
\nc{\CU}{{\mcl{U}}} \nc{\CUb}{{\wbar{\CU}}} \nc{\CUt}{{\wtd{\CU}}} \nc{\CUh}{{\wht{\CU}}}
\nc{\CV}{{\mcl{V}}} \nc{\CVb}{{\wbar{\CV}}} \nc{\CVt}{{\wtd{\CV}}} \nc{\CVh}{{\wht{\CV}}}
\nc{\CW}{{\mcl{W}}} \nc{\CWb}{{\wbar{\CW}}} \nc{\CWt}{{\wtd{\CW}}} \nc{\CWh}{{\wht{\CW}}}
\nc{\CX}{{\mcl{X}}} \nc{\CXb}{{\wbar{\CX}}} \nc{\CXt}{{\wtd{\CX}}} \nc{\CXh}{{\wht{\CX}}}
\nc{\CY}{{\mcl{Y}}} \nc{\CYb}{{\wbar{\CY}}} \nc{\CYt}{{\wtd{\CY}}} \nc{\CYh}{{\wht{\CY}}}
\nc{\CZ}{{\mcl{Z}}} \nc{\CZb}{{\wbar{\CZ}}} \nc{\CZt}{{\wtd{\CZ}}} \nc{\CZh}{{\wht{\CZ}}}

\let\eps\epsilon
\let\ups\upsilon
\let\veps\varepsilon
\let\vtht\vartheta

\let\vsgm\varsigma
\let\vphi\varphi
\let\vrho\varrho

\nc{\alphab}{{\bar{\alpha}}} \nc{\alphat}{{\td{\alpha}}} \nc{\alphah}{{\hat{\alpha}}}
\nc{\betab}{{\bar{\beta}}}   \nc{\betat}{{\td{\beta}}}   \nc{\betah}{{\hat{\beta}}} 
\nc{\gammab}{{\bar{\gamma}}} \nc{\gammat}{{\td{\gamma}}} \nc{\gammah}{{\hat{\gamma}}} 
\nc{\deltab}{{\bar{\delta}}} \nc{\deltat}{{\td{\delta}}} \nc{\deltah}{{\hat{\delta}}} 
\nc{\epsilonb}{{\bar{\eps}}} \nc{\epsilont}{{\td{\eps}}} \nc{\epsilonh}{{\hat{\eps}}} 
\nc{\vepsb}{{\bar{\veps}}}   \nc{\vepst}{{\td{\veps}}}   \nc{\vepsh}{{\hat{\veps}}} 
\nc{\zetab}{{\bar{\zeta}}}   \nc{\zetat}{{\td{\zeta}}}   \nc{\zetah}{{\hat{\zeta}}} 
\nc{\etab}{{\bar{\eta}}}     \nc{\etat}{{\td{\eta}}}     \nc{\etah}{{\hat{\eta}}} 
\nc{\thetab}{{\bar{\theta}}} \nc{\thetat}{{\td{\theta}}} \nc{\thetah}{{\hat{\theta}}} 
\nc{\vthetab}{{\bar{\vtht}}} \nc{\vthetat}{{\td{\vtht}}} \nc{\vthetah}{{\hat{\vtht}}} 
\nc{\lambdab}{{\bar{\lambda}}} \nc{\lambdat}{{\td{\lambda}}} \nc{\lambdah}{{\hat{\lambda}}} 
\nc{\iotab}{{\bar{\iota}}}   \nc{\iotat}{{\td{\iota}}}   \nc{\iotah}{{\hat{\iota}}} 
\nc{\kappab}{{\bar{\kappa}}} \nc{\kappat}{{\td{\kappa}}} \nc{\kappah}{{\hat{\kappa}}} 
\nc{\lmdb}{{\bar{\lmd}}}     \nc{\lmdt}{{\td{\lmd}}}     \nc{\lmdh}{{\hat{\lmd}}} 
\nc{\mub}{{\bar{\mu}}}       \nc{\mut}{{\td{\mu}}}       \nc{\muh}{{\hat{\mu}}} 
\nc{\nub}{{\bar{\nu}}}       \nc{\nut}{{\td{\nu}}}       \nc{\nuh}{{\hat{\nu}}} 
\nc{\xib}{{\bar{\xi}}}       \nc{\xit}{{\td{\xi}}}       \nc{\xih}{{\hat{\xi}}} 
\nc{\pib}{{\bar{\pi}}}       \nc{\pit}{{\td{\pi}}}       \nc{\pih}{{\hat{\pi}}} 
\nc{\vpib}{{\bar{\vpi}}}     \nc{\vpit}{{\td{\vpi}}}     \nc{\vpih}{{\hat{\vpi}}} 
\nc{\rhob}{{\bar{\rho}}}     \nc{\rhot}{{\td{\rho}}}     \nc{\rhoh}{{\hat{\rho}}} 
\nc{\vrhob}{{\bar{\vrho}}}   \nc{\vrhot}{{\td{\vrho}}}   \nc{\vrhoh}{{\hat{\vrho}}} 
\nc{\sigmab}{{\bar{\sigma}}} \nc{\sigmat}{{\td{\sigma}}} \nc{\sigmah}{{\hat{\sigma}}} 
\nc{\vsigmab}{{\bar{\vsgm}}} \nc{\vsigmat}{{\td{\vsgm}}} \nc{\vsigmah}{{\hat{\vsgm}}} 
\nc{\taub}{{\bar{\tau}}}     \nc{\taut}{{\td{\tau}}}     \nc{\tauh}{{\hat{\tau}}} 
\nc{\upsb}{{\bar{\ups}}} \nc{\upst}{{\td{\ups}}} \nc{\upsh}{{\hat{\ups}}} 
\nc{\phib}{{\bar{\phi}}}     \nc{\phit}{{\td{\phi}}}     \nc{\phih}{{\hat{\phi}}} 
\nc{\varphib}{{\bar{\vphi}}}   \nc{\varphit}{{\td{\vphi}}}   \nc{\varphih}{{\hat{\vphi}}} 
\nc{\chib}{{\bar{\chi}}}     \nc{\chit}{{\td{\chi}}}     \nc{\chih}{{\hat{\chi}}} 
\nc{\psib}{{\bar{\psi}}}     \nc{\psit}{{\td{\psi}}}     \nc{\psih}{{\hat{\psi}}} 
\nc{\omegab}{{\bar{\omega}}} \nc{\omegat}{{\td{\omega}}} \nc{\omegah}{{\hat{\omega}}} 

\nc{\Gammab}{{\wbar{\Gamma}}}     \nc{\Gammat}{{\wtd{\Gamma}}}     \nc{\Gammah}{{\wht{\Gamma}}}
\nc{\Deltab}{{\wbar{\Delta}}}     \nc{\Deltat}{{\wtd{\Delta}}}     \nc{\Deltah}{{\wht{\Delta}}}
\nc{\Thetab}{{\wbar{\Theta}}}     \nc{\Thetat}{{\wtd{\Theta}}}     \nc{\Thetah}{{\wht{\Theta}}}
\nc{\Lambdab}{{\wbar{\Lambda}}}   \nc{\Lambdat}{{\wtd{\Lambda}}}   \nc{\Lambdah}{{\wht{\Lambda}}}
\nc{\Xib}{{\wbar{\Xi}}}           \nc{\Xit}{{\wtd{\Xi}}}           \nc{\Xih}{{\wht{\Xi}}}
\nc{\Pib}{{\wbar{\Pi}}}           \nc{\Pit}{{\wtd{\Pi}}}           \nc{\Pih}{{\wht{\Pi}}}
\nc{\Sigmab}{{\wbar{\Sigma}}}     \nc{\Sigmat}{{\wtd{\Sigma}}}     \nc{\Sigmah}{{\wht{\Sigma}}}
\nc{\Upsilonb}{{\wbar{\Upsilon}}} \nc{\Upsilont}{{\wtd{\Upsilon}}} \nc{\Upsilonh}{{\wht{\Upsilon}}}
\nc{\Phib}{{\wbar{\Phi}}}         \nc{\Phit}{{\wtd{\Phi}}}         \nc{\Phih}{{\wht{\Phi}}}
\nc{\Psib}{{\wbar{\Psi}}}         \nc{\Psit}{{\wtd{\Psi}}}         \nc{\Psih}{{\wht{\Psi}}}
\nc{\Omegab}{{\wbar{\Omega}}}     \nc{\Omegat}{{\wtd{\Omega}}}     \nc{\Omegah}{{\wht{\Omega}}}

\newcommand{\rmd}{\mathrm{d}}

\newlength{\qsep}
\usetikzlibrary{decorations.markings, arrows, intersections}
\tikzset{
  x=\qsep, y=\qsep,  font=\smaller,
  ->-/.style={decoration={
      markings, mark=at position #1 with
      {\arrow{>}}},postaction={decorate}},
  -<-/.style={decoration={
      markings, mark=at position #1 with
      {\arrow{<}}},postaction={decorate}},
  node/.style={draw, fill=white, shape=circle, minimum size=7pt, inner
    sep=0pt},
  gnode/.style={node},
  dgnode/.style={node, densely dashed},
  ggnode/.style={node, double},
  fnode/.style={node, shape=rectangle},
  tnode/.style={fnode, double, minimum size=12pt},
  q-/.style={-},
  q->/.style={->,  >=stealth, shorten >=1pt, font=\smaller[2]},
  q<-/.style={q->, <-, shorten >=0pt, shorten <=1pt},
  eq-/.style={double, double distance=2pt},
}

\newcommand{\fnode}[1]{
  \mathbin{
    \mathchoice
    {\tikz[baseline=(x.base)]{\node[fnode] (x) at (0,0) {#1\vphantom{x}};}}
    {\tikz[baseline={([yshift=-0.7pt]x.base)}]{\node[fnode] (x) at (0,0) {#1\vphantom{x}};}}
    {\tikz[baseline=(x.base)]{\node[fnode, font=\scriptsize, minimum size=6pt] (x) at (0,0) {$#1$};}}
    {\tikz[baseline=(x.base)]{\node[fnode] (x) at (0,0) {$#1$};}}
}}

\newcommand{\gnode}[1]{
  \mathbin{
    \mathchoice
    {\tikz[baseline=(x.base)]{\node[gnode] (x) at (0,0) {#1\vphantom{x}};}}
    {\tikz[baseline={([yshift=-0.7pt]x.base)}]{\node[gnode] (x) at (0,0) {#1\vphantom{x}};}}
    {\tikz[baseline=(x.base)]{\node[gnode, font=\scriptsize] (x) at (0,0) {$#1$};}}
    {\tikz[baseline=(x.base)]{\node[gnode] (x) at (0,0) {$#1$};}}
}}

\newcommand{\ggnode}[1]{
  \mathbin{
    \mathchoice
    {\tikz[baseline=(x.base)]{\node[gnode, double] (x) at (0,0) {#1\vphantom{x}};}}
    {\tikz[baseline={([yshift=-0.7pt]x.base)}]{\node[gnode, double] (x) at (0,0) {#1\vphantom{x}};}}
    {\tikz[baseline=(x.base)]{\node[gnode, double, font=\scriptsize] (x) at (0,0) {$#1$};}}
    {\tikz[baseline=(x.base)]{\node[gnode, double] (x) at (0,0) {$#1$};}}
}}

\newcommand{\RM}{\check{R}}
\newcommand{\LM}{\check{L}}
\newcommand{\CRM}{\check{\CR}}

\tikzset{
  r-/.style={-, thick},
  r->/.style={r-, ->},
  r<-/.style={r-, <-},
  r->-/.style={->-=#1, thick},
}

\tikzset{
  z-/.style={-},
  z->/.style={z-, ->},
  z<-/.style={z->, <-},
  dr-/.style={-, densely dashed, thick},
  dr->/.style={dr-, ->},
  dz<-/.style={dr-, <-},
  dz<->/.style={dr-, <->},
  wz-/.style={z-, densely dotted}, 
  wz->/.style={wz-, ->},
  wz<-/.style={wz-, <-},
  shaded/.style={fill=black!20},
  lshaded/.style={fill=midnightblue!20},
  dshaded/.style={fill=midnightblue!50},
  frame/.style={draw=olive},
  ws/.style={fill=olive!5},
  boundary/.style={draw=maroon},
  fboundary/.style={boundary, double},
}

\tikzset{
  minp/.style={draw, shape=circle, minimum size=3pt, inner 
    sep=0pt, font=\tiny},
  maxp/.style={minp, double, double distance=1pt, fill=white, minimum
    size=5pt},
}
\newcommand{\V}{V}

\newcommand{\NS}[1]{\mathrm{NS#1}}
\newcommand{\D}[1]{\mathrm{D}#1}

\newcommand{\thy}{\mathsf{T}}
\newcommand{\latmod}{\mathsf{L}}
\newcommand{\iu}{\mathrm{i}}

\usepackage[super]{cite}
\usepackage{graphicx}
\begin{document}

\markboth{Junya Yagi}{Branes and Integrable Lattice Models}


\title{Branes and integrable lattice models}

\author{\footnotesize Junya Yagi}

\address{Faculty of Physics, University of Warsaw,
  ul.\ Pasteura 5, 02--093 Warsaw, Poland\\
junya.yagi@fuw.edu.pl}

\maketitle

\pub{Received (Day Month Year)}{Revised (Day Month Year)}

\begin{abstract}
  This is a brief review of my work on the correspondence between
  four-dimensional $\CN = 1$ supersymmetric field theories realized by
  brane tilings and two-dimensional integrable lattice models.  I
  explain how to construct integrable lattice models from extended
  operators in partially topological quantum field theories, and
  elucidate the correspondence as an application of this construction.

\end{abstract}

\ccode{PACS Nos.: 11.25.-w, 02.30.Ik, 05.50.+q}

\section{Introduction}

In supersymmetric field theories, exact computations are often
possible for a limited class of physical quantities.  Supersymmetric
indices are primary examples of such quantities, and have been
extensively studied in connection with gauge theory dualities,
holography and other interesting phenomena.

Around 2010, it was discovered that supersymmetric indices of certain
four-dimensional $\CN = 1$ supersymmetric gauge theories coincide with
the partition functions of two-dimensional integrable lattice models
in statistical mechanics.\cite{Bazhanov:2010kz, Spiridonov:2010em,
  Bazhanov:2011mz} As was later recognized\cite{Yamazaki:2012cp},
these gauge theories are realized by particular configurations of
branes in string theory, called brane tilings.\cite{Hanany:2005ve,
  Franco:2005rj} The lattice models in question are known as the
Bazhanov--Sergeev models\cite{Bazhanov:2010kz, Bazhanov:2011mz} and
have continuous spin variables.  The Yang--Baxter equations that
guarantee the integrability of the models are integral identities
obeyed by the elliptic gamma function.  On the gauge theory side, they
translate to the invariance of the indices under Seiberg
duality.\cite{Seiberg:1994pq}

This article provides a concise review of the main results of
Ref.~8
where the above correspondence was elucidated from the perspective of
topological quantum field theories (TQFTs).  Also discussed is the
role played by surface defects, which was partly understood in Ref.~9.
I hope that this review will serve as an introduction to the beautiful
yet largely unexplored connections between branes, supersymmetric
field theories, TQFTs and integrable lattice models.

\section{TQFTs with Extra Dimensions and Integrable Lattice Models}

Underlying the correspondence between brane tilings and integrable
lattice models is a general method to construct such models from
extended operators in partially topological quantum field theories.
To begin with, I present this construction at a formal level.  The
essential idea of it was developed by Costello.\cite{Costello:2013zra,
  Costello:2013sla}

\subsection{Lattice models from line operators in two-dimensional TQFTs}

\begin{figure}
  \centering
  \subfloat[\label{fig:lattice}]{
    \begin{tikzpicture}
      \fill[ws] (0,0) rectangle (3,2);

      \begin{scope}[shift={(0.75,0)}]
        \draw[r->] (0,0) -- (0,2);
        \draw[r->] (1,0) -- (1,2);
        \draw[r->] (2,0) -- (2,2);
      \end{scope}
      
      \begin{scope}[shift={(0,0.75)}]
        \draw[r->] (0,0) -- (3,0);
        \draw[r->] (0,1) -- (3,1);
      \end{scope}

      

      \draw[frame] (0,0) rectangle (3,2);
    \end{tikzpicture}
  }
  \quad
  \subfloat[\label{fig:lattice-holes}]{
    \begin{tikzpicture}
      \fill[ws] (0,0) rectangle (3,2);

      \begin{scope}[shift={(0.75,0)}]
        \draw[r->] (0,0) -- (0,2);
        \draw[r->] (1,0) -- (1,2);
        \draw[r->] (2,0) -- (2,2);
      \end{scope}
      
      \begin{scope}[shift={(0,0.75)}]
        \draw[r->] (0,0) -- (3,0);
        \draw[r->] (0,1) -- (3,1);
      \end{scope}

      \begin{scope}[shift={(0,0.25)}]
        \draw[boundary, densely dashed] (0,0) -- (3,0);
        \draw[boundary, densely dashed] (0,1) -- (3,1);
      \end{scope}
      
      \begin{scope}[shift={(0.25,0)}]
        \draw[boundary, densely dashed] (0,0) -- (0,2);
        \draw[boundary, densely dashed] (1,0) -- (1,2);
        \draw[boundary, densely dashed] (2,0) -- (2,2);
      \end{scope}

      \begin{scope}[shift={(0.25,0.25)}]
        \node[ggnode, boundary] at (0,0) {};
        \node[ggnode, boundary] at (1,0) {};
        \node[ggnode, boundary] at (2,0) {};
        \node[ggnode, boundary] at (0,1) {};
        \node[ggnode, boundary] at (1,1) {};
        \node[ggnode, boundary] at (2,1) {};
      \end{scope}

      \draw[frame] (0,0) rectangle (3,2);
    \end{tikzpicture}
  }
  \quad
  \subfloat[\label{fig:lattice-spins}]{
    \begin{tikzpicture}
      \fill[ws] (0,0) rectangle (3,2);

      \begin{scope}[shift={(0.75,0)}]
        \draw[r->] (0,0) -- (0,2);
        \draw[r->] (1,0) -- (1,2);
        \draw[r->] (2,0) -- (2,2);
      \end{scope}
      
      \begin{scope}[shift={(0,0.75)}]
        \draw[r->] (0,0) -- (3,0);
        \draw[r->] (0,1) -- (3,1);
      \end{scope}

      \begin{scope}[shift={(0.25,0.25)}]
        \node[ggnode] at (0,0) {};
        \node[ggnode] at (1,0) {};
        \node[ggnode] at (2,0) {};
        \node[ggnode] at (0,1) {};
        \node[ggnode] at (1,1) {};
        \node[ggnode] at (2,1) {};
      \end{scope}

      \begin{scope}[shift={(0.25,0.75)}] at
        \node[gnode] at (0,0) {};
        \node[gnode] at (1,0) {};
        \node[gnode] at (2,0) {};
        \node[gnode] at (0,1) {};
        \node[gnode] at (1,1) {};
        \node[gnode] at (2,1) {};
      \end{scope}

      \begin{scope}[shift={(0.75,0.25)}] at
        \node[gnode] at (0,0) {};
        \node[gnode] at (1,0) {};
        \node[gnode] at (2,0) {};
        \node[gnode] at (0,1) {};
        \node[gnode] at (1,1) {};
        \node[gnode] at (2,1) {};
      \end{scope}

      \draw[frame] (0,0) rectangle (3,2);
    \end{tikzpicture}
  }
  \caption{Construction of a lattice model from line operators. (a) A
    lattice of line operators on a torus. (b)~The torus with holes
    obtained by gluing of pieces. (c) The corresponding spin model.}
  \label{fig:2x3}
\end{figure}
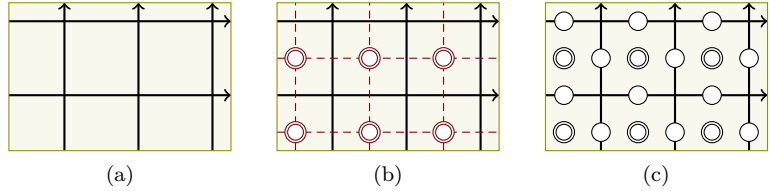

Suppose that we have a two-dimensional TQFT $\thy$ equipped with line
operators.  Place this theory on a torus $T^2$, and wrap line
operators $\CL_i$, $i = 1$, $\dotsc$, $l$ around closed curves $C_i$
in such a way that they form an $m \times n$ lattice.
Fig.~\ref{fig:lattice} illustrates the case $(m,n) = (2,3)$.  We wish
to compute the correlation function for this configuration of line
operators on the torus.

Our strategy is to break up the torus into small pieces, and first
perform the path integral piecewise.  Then we combine the results
from these pieces and reconstruct the original correlation function.%
\footnote{In the axiomatic language, we will compute the correlation
  function by embedding the closed TQFT with line operators into an
  open/closed TQFT with line operators.}

Consider the following piece of surface containing an intersection of
two line operators $\CL_i$ and $\CL_j$:
\begin{equation}
  \label{eq:piece}
  \begin{tikzpicture}[scale=0.5, baseline=(x.base)]
    \node (x) at (1,1) {\vphantom{x}};

    \fill[ws]
    (0.5,0) -- (1.5,0) arc (180:90:0.5)
    -- (2,1.5) arc (270:180:0.5)
    -- (0.5,2) arc (0:-90:0.5)
    -- (0,0.5) arc (90:0:0.5) -- cycle;

    \draw[r->] (0,1) node[left] {$i$} -- (2,1);
    \draw[r->] (1,0) node[below] {$j$} -- (1,2);
   
    \draw[boundary] (0.5,0) -- (1.5,0)
    (2,0.5) -- (2,1.5)
    (1.5,2) -- (0.5,2) 
    (0,1.5) -- (0,0.5);

    \draw[fboundary] (0.5,0) arc (0:90:0.5)
    node[midway, below left=-2pt] {$b$};
    \draw[fboundary] (2,0.5) arc (90:180:0.5)
    node[midway, below right=-2pt] {$c$};
    \draw[fboundary] (2,1.5) arc (-90:-180:0.5)
    node[midway, above right=-2pt] {$d$};
    \draw[fboundary] (0.5,2) arc (0:-90:0.5)
    node[midway, above left=-2pt] {$a$};
  \end{tikzpicture}
  \,.
\end{equation}
This picture represents the worldsheet of two scattering open strings,
each carrying a particle whose worldline is one of the line operators.
The end points of the strings sweep out the double-lined arcs.  The
strings are attached to D-branes there, and subject to boundary
conditions which are specified by labels $a$, $b$, $c$, $d$
(``Chan--Paton factors'').  We denote the set of boundary conditions
by $B$.

The path integral for the above surface produces a linear map
\begin{equation}
  \RM_{ij}\biggl(
  \begin{array}{cc}
    a & d \\
    b & c
  \end{array}
  \biggr)
  \colon
  \V_{ab,i} \otimes \V_{bc,j}
  \to
  \V_{ad,j} \otimes \V_{dc,i}
  \,,
\end{equation}
where $\V_{ab,i}$ is the space of states on an interval intersected
by~$\CL_i$, with the boundary conditions on the left and the right
ends being $a$ and $b$, respectively.  We call this map the
\emph{R-matrix} (or \emph{R-operator}) associated with this decorated
surface.

To reconstruct the lattice on the whole torus, we glue pieces similar
to the above together.  Gluing amounts to composing the corresponding
R-matrices.  For example, gluing two pieces horizontally gives
\begin{equation}
  \begin{tikzpicture}[scale=0.5, baseline={(x.base)}]
    \node (x) at (1,1) {\vphantom{x}};

    \fill[ws]
    (0.5,0) -- (1.5,0) arc (180:90:0.5)
    -- (2,1.5) arc (270:180:0.5)
    -- (0.5,2) arc (0:-90:0.5)
    -- (0,0.5) arc (90:0:0.5) -- cycle;

    \draw[r-] (0,1) node[left] {$i$} -- (2,1);
    \draw[r->] (1,0) node[below] {$j$} -- (1,2);

    \draw[boundary, densely dashed] (2,1.5) -- (2,0.5);

    \draw[boundary] (0.5,0) -- (1.5,0)
    (1.5,2) -- (0.5,2) 
    (0,1.5) -- (0,0.5);

    \begin{scope}[shift={(2,0)}]
      \fill[ws]
      (0.5,0) -- (1.5,0) arc (180:90:0.5)
      -- (2,1.5) arc (270:180:0.5)
      -- (0.5,2) arc (0:-90:0.5)
      -- (0,0.5) arc (90:0:0.5) -- cycle;

      \draw[r->] (0,1) -- (2,1);
      \draw[r->] (1,0) node[below] {$k$} -- (1,2);
   
      \draw[boundary]
      (0.5,0) -- (1.5,0)
      (2,0.5) -- (2,1.5)
      (1.5,2) -- (0.5,2);

      \draw[fboundary] (2,0.5) arc (90:180:0.5)
      node[midway, below right=-2pt] {$f$};
      \draw[fboundary] (2,1.5) arc (-90:-180:0.5)
      node[midway, above right=-2pt] {$e$};
    \end{scope}

    \draw[fboundary] (0.5,0) arc (0:90:0.5)
    node[midway, below left=-2pt] {$b$};
    \draw[fboundary] (2.5,0) arc (0:180:0.5)
    node[midway, below=0.75pt] {$c$};
    \draw[fboundary] (2.5,2) arc (0:-180:0.5)
    node[midway, above=0.75pt] {$d$};
    \draw[fboundary] (0.5,2) arc (0:-90:0.5)
    node[midway, above left=-2pt] {$a$};
  \end{tikzpicture}
  =
  \RM_{ik}\biggl(
  \begin{array}{cc}
    d & e \\
    c & f
  \end{array}
  \biggr)
  \circ_{V_{dc,i}}
  \RM_{ij}\biggl(
  \begin{array}{cc}
    a & d \\
    b & c
  \end{array}
  \biggr)
  \,.
\end{equation}
The torus thus obtained by gluing, however, has holes in it and looks
as in Fig.~\ref{fig:lattice-holes}.  On the boundaries of these holes
are imposed various boundary conditions, specified by labels $a$, $b$,
etc.  We must fill these holes.

This is achieved by summation over the boundary conditions.  The path
integral on a finite-length cylinder, with boundary condition $a$
imposed on one end, defines on the other end a closed string state
$\ket{a}$, called a boundary state.  Similarly, the path integral on a
disk with no insertion of operators defines a state $\ket{1}$ on the
boundary.  Assume that we have chosen the set $B$ to be sufficiently
large so that $\ket{1}$ can be written as a superposition of boundary
states:
\begin{equation}
  \label{eq:ket-1}
  \ket{1} = \sum_{a \in B} c_a \ket{a}
  \,.
\end{equation}
Then, summing over the boundary conditions we get $\ket{1}$ on the
boundary of each hole, which may in turn be replaced with a disk:
\begin{equation}
  \sum_{a \in B}
    c_a \biggl(\!%
    \begin{tikzpicture}[scale=0.18, baseline={(x.base)}]
      \node (x) at (1,0) {\vphantom{x}};

      \fill[ws] (0,-2) rectangle (4,2);

      \begin{scope}[shift={(-1.5,0)}]
        \clip (2.9,-2.1) rectangle (4,2.1);
        \draw[boundary, densely dotted] (4,0) ellipse (1 and 2);
      \end{scope}
      
      \begin{scope}[shift={(-1.5,0)}]
        \clip (4,-2.1) rectangle (5.1,2.1);
        \draw[boundary, densely dashed, ws] (4,0) ellipse (1 and 2);
      \end{scope}

      \begin{scope}
        \clip (4,-2.1) rectangle (5.1,2.1);
        \fill[ws] (4,0) ellipse (1 and 2);
      \end{scope}
      
      \draw (0,2) -- (4,2);
      \draw (0,-2) -- (4,-2);
      
      \draw[fboundary, fill=olive!10] (0,0) ellipse (1 and 1.9);
      \node[left] at (-1,0) {$a$};
    \end{tikzpicture}
    \biggr)
    =
    \sum_{a \in B}
    c_a \biggl(\!%
    \begin{tikzpicture}[scale=0.18, baseline={(x.base)}]
      \node (x) at (1,0) {\vphantom{x}};

      \fill[ws] (2.5,-2) rectangle (4,2);

      \begin{scope}[shift={(-1.5,0)}]
        \clip (2.9,-2.1) rectangle (4,2.1);
        \draw[boundary, fill=olive!10] (4,0) ellipse (1 and 2);
      \end{scope}
      
      \begin{scope}[shift={(-1.5,0)}]
        \clip (4,-2.1) rectangle (5.1,2.1);
        \draw[boundary, fill=olive!10] (4,0) ellipse (1 and 2);
      \end{scope}

      \begin{scope}
        \clip (4,-2.1) rectangle (5.1,2.1);
        \fill[ws] (4,0) ellipse (1 and 2);
      \end{scope}
      
      \draw (2.5,2) -- (4,2);
      \draw (2.5,-2) -- (4,-2);
      \node[left] at (1.5,0) {$\ket{a}$};
    \end{tikzpicture}%
    \biggr)
    =
    \begin{tikzpicture}[scale=0.18, baseline={(x.base)}]
      \node (x) at (1,0) {\vphantom{x}};

      \fill[ws] (2.5,-2) rectangle (4,2);

      \begin{scope}[shift={(-1.5,0)}]
        \clip (2.9,-2.1) rectangle (4,2.1);
        \draw[boundary, fill=olive!10] (4,0) ellipse (1 and 2);
      \end{scope}
      
      \begin{scope}[shift={(-1.5,0)}]
        \clip (4,-2.1) rectangle (5.1,2.1);
        \draw[boundary, fill=olive!10] (4,0) ellipse (1 and 2);
      \end{scope}

      \begin{scope}
        \clip (4,-2.1) rectangle (5.1,2.1);
        \fill[ws] (4,0) ellipse (1 and 2);
      \end{scope}
      
      \draw (2.5,2) -- (4,2);
      \draw (2.5,-2) -- (4,-2);
      \node[left] at (1.5,0) {$\ket{1}$};
    \end{tikzpicture}%
    =
    \begin{tikzpicture}[scale=0.18, baseline={(x.base)}]
      \node (x) at (2,0) {\vphantom{x}};

      \begin{scope}[shift={(1.5,0)}]
        \fill[ws] (2.5,-2) rectangle (4,2);

        \begin{scope}
          \clip (4,-2.1) rectangle (5.1,2.1);
          \fill[ws] (4,0) ellipse (1 and 2);
        \end{scope}
        
        \draw (2.5,2) -- (4,2);
        \draw (2.5,-2) -- (4,-2);
      \end{scope}
      
      \begin{scope}
        \clip (0.9,-2.1) rectangle (4,2.1);
        \draw[ws] (4,0) ellipse (2.5 and 2);
        \draw[boundary, densely dotted] (4,0) ellipse (1 and 2);
      \end{scope}
      
      \begin{scope}
        \clip
        (4,-2.1) rectangle (5.1,2.1);
        \draw[boundary, densely dashed, ws] (4,0) ellipse (1 and 2);
      \end{scope}
    \end{tikzpicture}
    \,.
\end{equation}
The holes are filled and disappear from the torus, as desired.


      
      
      
      


Having understood how to reconstruct the correlation function on the
torus, let us interpret this procedure as an operation in statistical
mechanics.  To this end, choose a basis for the open string state
space $\V_{ab,i}$ for each $a$, $b$ and $i$.  According to what we
have just found, the procedure for computing the correlation function
consists of three steps: First, pick a basis state for every side of
the pieces comprising the torus and a boundary condition for every
hole in the torus.  Second, take the product of the corresponding
R-matrix elements from all pieces as well as the coefficients $c_a$
associated with the boundary conditions from all holes.  Finally, sum
over all possible assignments of basis states and boundary conditions.

The first step may be alternatively thought of as assigning basis
states to the circles $\gnode{}$ and boundary conditions to the
double-lined circles $\ggnode{}$ on the torus shown in
Fig.~\ref{fig:lattice-spins}.  Rephrased in this way, it is clear that
the above procedure defines the partition function of a spin model.
The model has spins located at two kinds of sites, $\gnode{}$ and
$\ggnode{}$\,.  A spin at $\gnode{}$ takes values in the chosen basis
for the relevant open string state space, while that at $\ggnode{}$ is
valued in $B$.  The Boltzmann weights for their interactions are
determined by the R-matrix elements and the coefficients $c_a$.

Thus, we conclude that the correlation function for a lattice of line
operators coincides with the partition function of a spin model
defined on the same lattice:
\begin{equation}
  \biggvev{\prod_{i=1}^l \CL_i(C_i)}_{\thy, T^2}
  =
  Z_{\latmod(\thy), \{\CL_i(C_i)\}}
  \,.
\end{equation}
Here, $\latmod(\thy)$ denotes the lattice model arising from the TQFT
$\thy$ by this construction and $\{\CL_i(C_i)\}$ is the lattice
formed by line operators $\CL_i$ wrapped around $C_i$.

If $B$ consists of a single boundary condition $a$, we simply write
$V_i$ for $V_{aa,i}$ and represent the R-matrix
$\RM_{ij}\colon \V_i \otimes \V_j \to \V_j \otimes \V_i$ by a crossing
of two lines:
\begin{equation}
  \RM_{ij}
  =
  \begin{tikzpicture}[scale=0.5, baseline={(x.base)}]
    \node (x) at (1,1) {\vphantom{x}};
    \draw[r->] (0,1) node[left] {$i$} -- (2,1);
    \draw[r->] (1,0) node[below] {$j$} -- (1,2);
  \end{tikzpicture}
  \,.
\end{equation}
In this case we can ignore the spins at $\ggnode{}$ since there is no
summation for them.  This means that $\latmod(\thy)$ is a \emph{vertex
  model}: the spins live on the edges of the lattice and interact at
the vertices.  We may think of $V_i$ as a vector space carried by
$\CL_i$.

If $\dim \V_{ab,i} = 1$ for all $a$, $b$ and $i$, we only sum over the
boundary conditions instead.  In this case the spins at $\gnode{}$ can
be ignored and $\latmod(\thy)$ is an \emph{interaction-round-a-face
  model} (or \emph{IRF model} for short): the spins are placed on the
faces and the interaction takes place among four spins surrounding a
vertex.

Formally, we can always recast our lattice model into a vertex model by
setting $\V_i = \bigoplus_{a,b \in B} \V_{ab,i}$ and declaring that
all newly introduced R-matrix elements, which correspond to scattering
processes with inconsistent Chan--Paton factors, vanish.  We can also
absorb the coefficients $c_a$ into the R-matrix elements by
appropriate rescaling.  In what follows this reformulation is
implicitly performed.

\subsection{Integrability from extra dimensions}

A remarkable aspect of this construction of lattice models is that it
allows us to understand integrability from a higher-dimensional point
of view.  This is the crucial observation by
Costello.\cite{Costello:2013zra, Costello:2013sla}

In our lattice model, consider a row where a horizontal line operator
$\CL_i$ intersects the vertical line operators $\CL_j$, $j = 1$,
$\dotsc$, $n$.  Concatenating the R-matrices in this row, we get the
row-to-row \emph{transfer matrix}
\begin{equation}
  T_i
  =
  \begin{tikzpicture}[scale=0.5, baseline=(x.base)]
    \node (x) at (1,0) {\vphantom{x}};
    
    \draw[r->, right hook->] (-0.1,0) node[left] {$i$} -- (1.1,0);
    \draw[r->, >=left hook] (1,0) -- (4.1,0);
    \node[fill=white, inner sep=1pt] at (2.5,0) {$\,\dots$};
    
    \begin{scope}[shift={(0.5,-0.5)}]
      \draw[r->] (0,0) node[below] {$1$} -- (0,1.2);
      \draw[r->] (1,0) node[below] {$2$} -- (1,1.2);
      \draw[r->] (3,0) node[below] {$n$} -- (3,1.2);
    \end{scope}
  \end{tikzpicture}
  = 
  \Tr_{\V_i}\bigl(
  \RM_{in} \circ_{\V_i} \dotsb \circ_{\V_i} \RM_{i1}
  \bigr)
  \,.
\end{equation}
(The hooks on the horizontal line are to remind us that the periodic
boundary condition is imposed.)  This object is an endomorphism of
$\bigotimes_{j=1}^n \V_j$ which maps a state just below $\CL_i$ to
another state just above it.  In terms of transfer matrices the
partition function is written as a trace:
\begin{equation}
  Z_{\latmod(\thy), \{\CL_i(C_i)\}}
  =
  \Tr\bigl(T_{n+m} \dotsb T_{n+1}\bigr)
  \,.
\end{equation}
In the TQFT context, $T_i$ may be regarded as a time-evolution
operator induced by~$\CL_i$, acting on the Hilbert space
$\bigotimes_{j=1}^n \V_j$.  Since the theory is topological, a state
evolves trivially unless it hits something -- line operators in the
present case.

Now, suppose that each line operator depends on a parameter which is
an element of some set $S$.  This parameter is called the
\emph{spectral parameter} of the lattice model.  We denote the
spectral parameter of $\CL_i$ by $u_i$.  Thus, $\RM_{ij}$ is a
function of two parameters $u_i$, $u_j$, whereas $T_i$ carries
$n+1$ parameters $u_1$, $\dotsc$, $u_n$ and $u_i$.  To avoid clutter,
we fix $u_1$, $\dotsc$, $u_n$ and suppress them below.

A vertex model is said to be \emph{integrable} if $T_i(u_i)$ is a
smooth function of $u_i$ (hence $S$ is a smooth manifold),
and moreover the relation
\begin{equation}
  \label{eq:[T,T]=0}
  \begin{tikzpicture}[scale=0.5, baseline=(x.base)]
    \node (x) at (1,0.5) {\vphantom{x}};
    
    \begin{scope}[shift={(0,0)}]
      \draw[r->, right hook->] (-0.1,0)
      node[left] {$i$} -- (1.1,0);
      \draw[r->, >=left hook] (1,0) -- (4.1,0);
      \node[fill=white, inner sep=1pt] at (2.5,0) {$\,\dots$};
    \end{scope}
    
    \begin{scope}[shift={(0,1)}]
      \draw[r->, right hook->] (-0.1,0)
      node[left] {$j$} -- (1.1,0);
      \draw[r->, >=left hook] (1,0) -- (4.1,0);
      \node[fill=white, inner sep=1pt] at (2.5,0) {$\,\dots$};
    \end{scope}
    
    \begin{scope}[shift={(0.5,-0.5)}]
      \draw[r->] (0,0) -- (0,2.2);
      \draw[r->] (1,0) -- (1,2.2);
      \draw[r->] (3,0) -- (3,2.2);
    \end{scope}
  \end{tikzpicture}
  \ =
  \begin{tikzpicture}[scale=0.5, baseline=(x.base)] 
   \node (x) at (1,0.5) {\vphantom{x}};
    
    \begin{scope}[shift={(0,0)}]
      \draw[r->, right hook->] (-0.1,0)
      node[left] {$j$} -- (1.1,0);
      \draw[r->, >=left hook] (1,0) -- (4.1,0);
      \node[fill=white, inner sep=1pt] at (2.5,0) {$\,\dots$};
    \end{scope}
    
    \begin{scope}[shift={(0,1)}]
      \draw[r->, right hook->] (-0.1,0)
      node[left] {$i$} -- (1.1,0);
      \draw[r->, >=left hook] (1,0) -- (4.1,0);
      \node[fill=white, inner sep=1pt] at (2.5,0) {$\,\dots$};
    \end{scope}
    
    \begin{scope}[shift={(0.5,-0.5)}]
      \draw[r->] (0,0) -- (0,2.2);
      \draw[r->] (1,0) -- (1,2.2);
      \draw[r->] (3,0) -- (3,2.2);
    \end{scope}
  \end{tikzpicture}
  \iff
  [T_i(u_i), T_j(u_j)] = 0
\end{equation}
holds for $u_i \neq u_j$.  When the model is integrable, we can find a
series of mutually commuting operators on $\bigotimes_{j=1}^n \V_j$
from the Taylor expansions of transfer matrices.

These conditions for integrability are naturally satisfied if the TQFT
has ``extra dimensions.''  In this scenario, we really start with a
higher-dimensional theory $\thy$ formulated on $S \times T^2$ that is
topological on $T^2$ but not on $S$.  We wrap line operators~$\CL_i$
around closed curves $u_i \times C_i$, where $u_i$ are points in $S$.
They may or may not have parameters.

To someone unaware of the presence of the extra dimensions $S$, the
theory appears as a two-dimensional TQFT, which we call $\thy[S]$.%
\footnote{Note that $\thy[S]$ is not the dimensional reduction of
  $\thy$ on $S$.  Here we are keeping all Kaluza--Klein modes and
  therefore $\thy[S]$ captures the full content of $\thy$.}
This observer finds that line operators $\CL_i[u_i]$ carrying
continuous parameters $u_i$ are wrapped around $C_i$ in the seemingly
two-dimensional spacetime $T^2$, and the correlation function for this
configuration is given by the partition function of a lattice model
$\latmod(\thy[S])$ defined on the lattice $\{\CL_{i}[u_i](C_i)\}$.
For a generic choice of the points $u_i$, the transfer matrices of
$\latmod(\thy[S])$ commute since the two horizontal line operators
in~\eqref{eq:[T,T]=0} may move freely and interchange their positions
due to the topological invariance along $T^2$; no phase transition
occurs when they pass each other as they do not meet in the full
spacetime $S \times T^2$.  Thus, the integrability follows from the
existence of extra dimensions, whose coordinates provide continuous
spectral parameters.

In fact, we can say more.  By the same logic, we deduce that
the \emph{unitarity relation}
\begin{equation}
  \begin{tikzpicture}[scale=0.9, baseline=(x.base)]
    \node (x) at (1,0.25) {\vphantom{x}};
    
    \draw[r->] (0,0) node[left] {$j$}
    to[out=0, in=180] (1,0.5) to[out=0, in=180] (2,0);
    
    \draw[r->] (0,0.5) node[left] {$i$}
    to[out=0, in=180] (1,0) to[out=0, in=180] (2,0.5);
  \end{tikzpicture}
  \ =
  \begin{tikzpicture}[scale=0.9, baseline=(x.base)]
    \node (x) at (1,0.25) {\vphantom{x}};

    \draw[r->] (0,0)  node[left] {$j$} -- (2,0);
    \draw[r->] (0,0.5) node[left] {$i$} -- (2,0.5);
  \end{tikzpicture}
  \iff
  \RM_{ji} \RM_{ij} = \id_{V_i \otimes V_j}
\end{equation}
and the \emph{Yang--Baxter equation}
\begin{equation}
  \label{eq:RRR}
  \begin{tikzpicture}[scale=0.5, baseline=(x.base)]
    \node (x) at (30:2) {\vphantom{x}};
    
    \draw[r->] (0,0) node[left] {$j$} -- ++(30:3);
    \draw[r->] (0,2) node[left] {$i$} -- ++(-30:3);
    \draw[r->] (-30:1) node[below] {$k$} -- ++(0,3);
  \end{tikzpicture}
  \ =
  \begin{tikzpicture}[scale=0.5, baseline=(x.base)]
    \node (x) at (30:1) {\vphantom{x}};
    
    \draw[r->] (0,0) node[left] {$j$} -- ++(30:3);
    \draw[r->] (0,1) node[left] {$i$} -- ++(-30:3);
    \draw[r->] (-30:2) node[below] {$k$} -- ++(0,3);
  \end{tikzpicture}
  \iff
  \begin{aligned}
    & (\id_{V_k} \otimes \RM_{ij}) (\RM_{ik} \otimes \id_{V_j}) (\id_{V_i} \otimes \RM_{jk})
    \\
    &\qquad
    =
    (\RM_{jk} \otimes \id_{V_i}) (\id_{V_j} \otimes \RM_{ik}) (\RM_{ij} \otimes \id_{V_k})
  \end{aligned}
\end{equation}
also hold.  (For brevity the spectral parameters are omitted.)  These
relations imply the commutativity of transfer matrices and hence the
integrability of the model.

\subsection{Correspondence with extended operators in extra
  dimensions}

The above argument generalizes in a couple of ways.  First of all, we
can formulate the higher-dimensional theory on a manifold of the form
$S \times \Sigma$, with $\Sigma$ being any surface, and put line
operators along various curves $u_i \times C_i$ in such a way that no
three curves intersect at a point on $\Sigma$.  In this situation we
get a spin model placed on the lattice drawn on $\Sigma$ by the curves
$C_i$.%
\footnote{If necessary, we introduce ``invisible'' line operators so
  that a lattice is formed.  They are the unit of the algebra of line
  operators and have no effect on the correlation function.}
The model is integrable in the sense that its R-matrix satisfies the
Yang--Baxter equation with spectral parameter.

The surface $\Sigma$ may have a boundary.  If it does, we make
``dents'' on the boundary between line operators and impose boundary
conditions there:
\begin{equation}
  \begin{tikzpicture}
    \fill[ws] (-0.3,0) rectangle (2.3,0.4);

    \draw[r-] (0,0) -- (0,0.4);
    \draw[r-] (1,0) -- (1,0.4);
    \draw[r-] (2,0) -- (2,0.4);

    \draw[boundary] (-0.3,0) -- (2.3,0);
  \end{tikzpicture}
  \ \to \
  \begin{tikzpicture}
    \fill[ws]
    (-0.3,0) -- (0.35,0) arc (180:0:0.15)
    -- (1.35,0) arc (180:0:0.15)
    -- (2.3,0) -- (2.3,0.4) -- (-0.3,0.4) -- cycle;

    \draw[r-] (0,0) -- (0,0.4);
    \draw[r-] (1,0) -- (1,0.4);
    \draw[r-] (2,0) -- (2,0.4);

    \draw[fboundary] (0.35,0) arc (180:0:0.15);
    \draw[fboundary] (1.35,0) arc (180:0:0.15);

    \draw[boundary]
    (-0.3,0) -- (0.35,0)
    (0.65,0) -- (1.35,0)
    (1.65,0) -- (2.3,0);
  \end{tikzpicture}
  \ .
\end{equation}
This process assigns spins to the faces touching the boundary.  These
spins, together with those assigned to the edges intersecting the
boundary, provide the data defining the TQFT states living on the
boundary (with dents).

Second, the line operators may descend from extended operators of
dimension greater than one.  Consider a theory $\thy$ formulated on
$S \times M \times \Sigma$, where $M$ is some manifold.  Suppose that
it is topological on $\Sigma$ and has extended operators~$\CE_i$ whose
codimension is greater than $\dim S$.  Place $\CE_i$ on submanifolds
of the form $u_i \times N _i \times C_i$.  Since $\thy[S \times M]$ --
the theory $\thy$ ``compactified'' on $S \times M$ and regarded as a
two-dimensional theory, though neither $S$ nor $M$ needs to be compact
-- is a TQFT, the correlation function of this configuration still
equals the partition function of an integrable lattice model
$\latmod(\thy[S \times M])$.  The model is placed on the lattice
constructed from the line operators $\CE_i[u_i \times N_i]$, the
images of $\CE_i$ in $\thy[S \times M]$.

In the previous paragraph we regarded our theory as a TQFT on
$\Sigma$, but we may also view it as a theory $\thy[\Sigma]$ on
$S \times M$.  In the latter theory $\CE_i$ appear as extended
operators $\CE_i[C_i]$ supported on $u_i \times N_i$, and we have
\begin{equation}
  \biggvev{
    \prod_{i=1}^l \CE_i[C_i](u_i \times N_i)
  }_{\thy[\Sigma], S \times M}
  =
  Z_{\latmod(\thy[S \times M]), \{\CE_i[u_i \times N_i](C_i)\}}
  \,.
\end{equation}
Thus we have arrived at a correspondence between the theory on the
extra dimensions $S \times M$ and the integrable lattice model on
$\Sigma$.  If $\Sigma$ has a boundary, this is an equality between
linear functionals on the Hilbert space of states.

\subsection{Higher-dimensional lattice models}

Our construction can also be extended to higher-dimensional lattice
models.  In a $d$-dimensional TQFT, a generic configuration of
$(d-1)$-dimensional extended operators makes a lattice.  The
correlation function for this configuration gives the partition
function of a $d$-dimensional lattice model.  If the theory has extra
dimensions and there is enough room there for these operators to avoid
one another, the model is integrable and satisfies a $d$-dimensional
analog of the Yang--Baxter equation.  For $d = 3$, the relevant
equation is Zamolodchikov's tetrahedron
equation.\cite{MR611994,Zamolodchikov:1981kf} Our argument shows that
the partition function equals a correlation function of extended
operators in a theory formulated on the extra dimensions.

\section{Branes and Integrable Lattice Models}

We have seen above that a lattice model is realized by a lattice of
line operators in a two-dimensional TQFT, and it is integrable if the
TQFT is embedded in higher dimensions and the line operators come from
extended operators localized in some directions of the extra
dimensions.  Now I explain how to get such structures of TQFTs with
extra dimensions using branes in string theory.

\begin{figure}
  \centering
  \begin{tikzpicture}[scale=0.6, align at top]
    \draw[ws, shift={(0.16,0.16)}] (0,0) rectangle (3,2);
    \draw[ws, shift={(0.08,0.08)}] (0,0) rectangle (3,2);
    \draw[ws] (0,0) rectangle (3,2);
    \node[below left] at (0,2) {NS5};
    \node[above left] at (0.5,-1) {$\D{p}_i$};

    \draw[fill=red!10] (1.5,0) -- ++(90:2) -- ++ (-135:1.5) 
    -- ++(-90:2) -- cycle;

    \draw[densely dotted] (1.5,0) -- ++(180:{1.5/sqrt(2)} );
  \end{tikzpicture}
  \quad
  \begin{tikzpicture}[scale=0.6, align at top]
    \draw[draw=none, shift={(0.16,0.16)}] (0,0) rectangle (3,2);
    \fill[ws] (0,0) rectangle (3,2);
    \draw[z-] (1.5,0) node[below] {$C_i$} -- (1.5,2);
    \node[below right] at (0,2) {$\Sigma$};
    \draw[frame] (0,0) rectangle (3,2);
  \end{tikzpicture}

  \caption{The D$p$-brane $\D{p}_i$ ending on the NS-branes creates a
    defect $\CE_{\D{p}_i}$ along $C_i$.}
  \label{fig:Dp-NS5}
\end{figure}
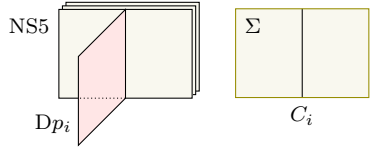

Consider a stack of $N$ NS5-branes supported on
$\R^{3,1} \times \Sigma \times 0$ in type II string theory in the
spacetime $\R^{3,1} \times T^*\Sigma \times \R^2$.  Here $\Sigma$ is a
surface (without boundary, for simplicity) embedded in $T^*\Sigma$ as
the zero section.  To this configuration, we introduce D$p$-branes
$\D{p}_i$ ending on the NS5-branes along curves $C_i$ on $\Sigma$, as
in Fig.~\ref{fig:Dp-NS5}.  Let their worldvolumes be
$\R^{p-2,1} \times \Sigma_i \times 0$, where $\R^{p-2,1}$ is a
subspace of $\R^{3,1}$ and $\Sigma_i$ are surfaces in $T^*\Sigma$ such
that $\del \Sigma_i \cap \Sigma = C_i$.  Provided that $\Sigma_i$ are
suitably chosen, this brane system preserves four supercharges.

The low-energy dynamics of the NS5-branes is governed by a
six-dimensional theory $\thy_\NS5$, which is either $\CN = (2,0)$
superconformal field theory of type $A_{N-1}$ or $\CN = (1,1)$ super
Yang--Mills theory with gauge group $\SU(N)$, depending on whether we
are considering type IIA or IIB string theory.  The theory $\thy_\NS5$
is formulated on $\R^{3,1} \times \Sigma$, with topological twist
along $\Sigma$ which breaks half of the sixteen supercharges.  In this
twisted theory, $\D{p}_i$ create $p$-dimensional defects
$\CE_{\D{p}_i}$ on $\R^{p-2,1} \times C_i$, reducing the number of
unbroken supercharges to four.  From the point of view of a
four-dimensional observer, this brane configuration gives half-BPS
defects $\CE_{\D{p}_i}[C_i]$ in an $\CN = 2$ theory
$\thy_\NS5[\Sigma]$.  The total system is invariant under a $\U(1)$
R-symmetry originating from the rotational symmetry on the $\R^2$
factor of the ten-dimensional spacetime.

Let us take a three-manifold $M$ and $(p-2)$-submanifolds $N_i$ of $M$,
and modify the above construction so that the worldvolumes of the
NS5-branes and the D$p$-branes become $S^1 \times M \times \Sigma$ and
$S^1 \times N_i \times \Sigma_i$, respectively.  At low energies, we
get the same theory $\thy_{\NS5}$ formulated on
$S^1 \times M \times \Sigma$, with $\CE_{\D{p}_i}$ located on
$S^1 \times N_i \times C_i$.  In general, this modification completely
breaks supersymmetry.  However, for certain choices of $M$ and $N_i$,
there is a string background in which a fraction of supersymmetry is
still preserved.  In such a background, the path integral computes the
\emph{supersymmetric index} of $\thy_{\NS5}$, defined with respect to
the Hilbert space on~$M \times \Sigma$ in the presence of the defects
$\CE_{\D{p}_i}$ inserted on $N_i \times C_i$.

A salient feature of supersymmetric indices is that they are protected
against continuous changes of various parameters of the theory.  This
is because a supersymmetric index is the trace of $(-1)^F$ over the
space of states annihilated by a set of supercharges, usually refined
by gradings with respect to some conserved charges commuting with
those supercharges.  Under variations of continuous parameters, such
states are created or annihilated in boson--fermion pairs and there is
no net change in the index.

For the same reason, the index of our theory is invariant under
deformations of the geometric data of $\Sigma$ and $C_i$, namely the
metric on $\Sigma$ and the shapes of $C_i$.  In other words,
$\thy_{\NS5}$ on $S^1 \times M \times \Sigma$ is topological on
$\Sigma$, as far as the computation of the index is concerned.


To connect the present setup to the one considered in the previous
section, we apply T-duality along $S^1$.  It turns $\D{p}_i$ into
D$(p-1)$-branes $\D{(p-1)}_i$ localized at points $u_i$ on the dual
circle $\St^1$, while sending the NS5-branes to those in the other
type II string theory.%
\footnote{More precisely, we obtain branes in an exotic variant of
  type II string theory with Euclidean D-branes, as we have applied
  timelike T-duality~\cite{Hull:1998vg}.}
The new NS5-branes produce the dual six-dimensional
theory~$\widetilde\thy_\NS5$ on $\St^1 \times M \times \Sigma$, and in
this theory $\D{(p-1)}_i$ create $(p-1)$-dimensional
defects~$\CE_{\D{(p-1)}_i}$ on~$u_i \times N_i \times C_i$.

Furthermore, we know that $\widetilde\thy_\NS5$ is topological
on~$\Sigma$ if we restrict the allowed operators to a subset which
includes these defects.  Thus we are in the situation studied in the
last section, and the correlation function for this configuration
coincides with the partition function of an integrable lattice model:
\begin{equation}
  \biggvev{
    \prod_{i=1}^l \CE_{\D{p}_i}[C_i](S^1 \times N_i) 
  }_{\thy_\NS5[\Sigma], S^1 \times M}
  =
  Z_{\latmod(\widetilde\thy_\NS5[\St^1 \times M]),
    \{\CE_{\D{(p-1)}_i}[u_i \times N_i](C_i)\}}
  \,.
\end{equation}
Here the correlation function is expressed in the original frame; it
implicitly depends on each spectral parameter $u_i$ through the
holonomy $\exp(2\pi\iu u_i)$ around $S^1$ of the gauge field for the
flavor symmetry $\U(1)_i$ supported on~$\D{p}_i$.  The holonomy
appears in the index as a refinement parameter, or \emph{fugacity},
associated with $\U(1)_i$.  In that context, $u_i$ is a \emph{chemical
  potential} for $\U(1)_i$.

\section{Integrable Lattice Models from Brane Tilings}

Finally, we apply the framework developed in the previous sections to
the main theme of this article: the integrable lattice models arising
from brane tilings.

\subsection{Brane tilings}

Let us consider the case $p = 5$ in the brane construction described
in the last section.  To conform with the standard convention, we go
to the S-dual frame where the D5-branes and the NS5-branes are
interchanged.  Thus, we have a stack of $N$ D5-branes wrapping
$S^1 \times M \times \Sigma$, together with NS5-branes $\NS5_i$
supported on $S^1 \times N_i \times \Sigma_i$ creating defects
$\CE_{\NS5_i}$ on $S^1 \times N_i \times C_i$ in the theory
$\thy_{\D5}$ on the D5-branes. In addition, we allow $\Sigma$ to have
a boundary where the 5-branes end on 7-branes.

The first thing to notice is that we necessarily have $N_i = M$, i.e.\
$\CE_{\NS5_i}$ wrap the whole $M$.  Accordingly, the half-BPS defects
$\CE_{\NS5_i}[C_i]$ in the four-dimensional $\CN = 2$ theory
$\thy_{\D5}[\Sigma]$ cover the entire spacetime $S^1 \times M$, and
may be thought of as changing $\thy_{\D5}[\Sigma]$ to a different
theory with $\CN = 1$ supersymmetry.

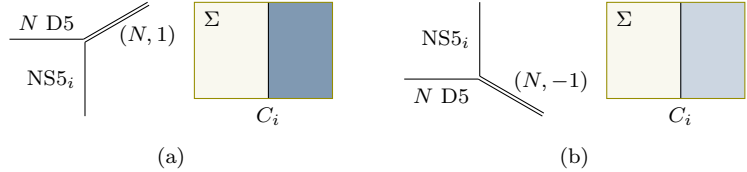
\begin{figure}
  \centering
  \subfloat[\label{fig:5BW-a}]{
    \begin{tikzpicture}[align at top]
      \begin{scope}[shift={(-0.02,0.04)}]
        \draw (0,0) -- node[above] {$N$ D5} (1,0);
        \draw (1,0) -- ++(30:1);
      \end{scope}

      \draw (1,-1) -- node[left] {$\NS5_i$} (1,0);
      \draw (1,0) -- node[below right=-4pt] {$(N,1)$} ++(30:1);
    \end{tikzpicture}
    \begin{tikzpicture}[align at top]
      \fill[ws] (0,0) rectangle ({1+sqrt(3)/2},1.3);
      \fill[dshaded] (1,0) rectangle ({1+sqrt(3)/2},1.3);
      \draw[z-] (1,0) node[below] {$C_i$} -- (1,1.3);
      \node[below right] at (0,1.3) {$\Sigma$};
      \draw[frame] (0,0) rectangle ({1+sqrt(3)/2},1.3);
    \end{tikzpicture}
  }
  \qquad
  \subfloat[\label{fig:5BW-b}]{
    \begin{tikzpicture}[align at top]
      \draw (0,0) -- node[below] {$N$ D5} (1,0);
      \draw (1,0) -- ++(-30:1);

      \begin{scope}[shift={(0.02,0.04)}]
        \draw (1,0) -- node[left] {$\NS5_i$} (1,1);
        \draw (1,0) -- node[above right=-4pt] {$(N,-1)$} ++(-30:1);
      \end{scope}
    \end{tikzpicture}
    \begin{tikzpicture}[align at top]
      \fill[ws] (0,0) rectangle ({1+sqrt(3)/2},1.3);
      \fill[lshaded] (1,0) rectangle ({1+sqrt(3)/2},1.3);
      \draw[z-] (1,0) node[below] {$C_i$} -- (1,1.3);
      \node[below right] at (0,1.3) {$\Sigma$};
      \draw[frame] (0,0) rectangle ({1+sqrt(3)/2},1.3);
    \end{tikzpicture}
  }
  \caption{An NS5-brane combines with a stack of $N$ D5-branes, forming
    (a) an $(N,1)$ 5-brane or (b) an $(N,-1)$ 5-brane. The 5-brane
    junction is a domain wall in $\thy_{\D{5}}$.  The shaded regions
    shown above support a nonzero NS5-brane charge $q = \pm1$.}
  \label{fig:5BW}
\end{figure}

Another peculiarity is that the NS5-branes cannot just end on the
D5-branes.  Rather, when an NS5-brane meets $N$ D5-branes, they
combine to form a bound state.  In the language of $(p,q)$ 5-branes,
this bound state is either an $(N,1)$ 5-brane or an $(N,-1)$ 5-brane,
depending on the relative positions of the branes; see
Fig.~\ref{fig:5BW}.  Therefore, $\CE_{\NS5_i}$ are domain walls in
$\thy_{\D5}$ partitioning the spacetime into regions with different
values of the NS5-brane charge~$q$.  (In this sense, $\thy_{\D5}$ is
not supported solely on D5-branes.)  The curves $C_i$ along which
these domain walls are located are known as \emph{zigzag paths}.
Across a zigzag path the value of $q$ jumps by one.

Conversely, given a configuration of curves $C_i$ on $\Sigma$ and a
5-brane charge assignment consistent with it, we can construct a
5-brane system whose zigzag paths are $C_i$: we take NS5-branes
approaching the D5-branes from transverse directions, and let them
meet along $C_i$ and form bound states over regions with $q \neq 0$.
Such a 5-brane system is called a \emph{brane
  tiling}\cite{Hanany:2005ve, Franco:2005rj} on $\Sigma$.  The reader
is referred to Refs.~16 and 17
for reviews of brane tilings.

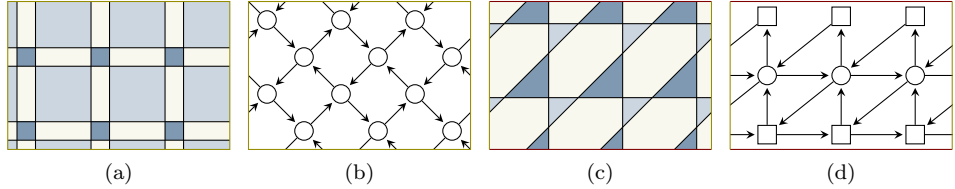
\begin{figure}
  \centering
  \subfloat[\label{fig:BT-D}]{
    \begin{tikzpicture}
      \fill[lshaded] (0,0) rectangle (3, 2);

      \begin{scope}[shift={(0.125,0)}]
        \fill[ws] (0,0) rectangle (0.25,2);
        \fill[ws] (1,0) rectangle (1.25,2);
        \fill[ws] (2,0) rectangle (2.25,2);
      \end{scope}

      \begin{scope}[shift={(0,0.125)}]
        \fill[ws] (0,0) rectangle (3,0.25);
        \fill[ws] (0,1) rectangle (3,1.25);
      \end{scope}

      \begin{scope}[shift={(0.125,0.125)}]
        \fill[dshaded] (0,0) rectangle (0.25,0.25);
        \fill[dshaded] (1,0) rectangle (1.25,0.25);
        \fill[dshaded] (2,0) rectangle (2.25,0.25);
        \fill[dshaded] (0,1) rectangle (0.25,1.25);
        \fill[dshaded] (1,1) rectangle (1.25,1.25);
        \fill[dshaded] (2,1) rectangle (2.25,1.25);
      \end{scope}

      \begin{scope}[shift={(0.125,0)}]
        \draw[z-] (0,0) -- (0,2);
        \draw[z-] (0.25,0) -- (0.25,2);
        \draw[z-] (1,0) -- (1,2);
        \draw[z-] (1.25,0) -- (1.25,2);
        \draw[z-] (2,0) -- (2,2);
        \draw[z-] (2.25,0) -- (2.25,2);
      \end{scope}

      \begin{scope}[shift={(0,0.125)}]
        \draw[z-] (0,0) -- (3,0);
        \draw[z-] (0,0.25) -- (3,0.25);
        \draw[z-] (0,1) -- (3,1);
        \draw[z-] (0,1.25) -- (3,1.25);
      \end{scope}

      \draw[frame] (0,0) rectangle (3, 2);
    \end{tikzpicture}
  }
  \subfloat[\label{fig:BT-D-quiver}]{
    \begin{tikzpicture}
      \begin{scope}[shift={(0.25, 0.25)}]
        \node[gnode] (i1) at (0.5, 0) {};
        \node[gnode] (i2) at (1.5, 0) {};
        \node[gnode] (i3) at (2.5, 0) {};
        \node[gnode] (j1) at (0.5, 1) {};
        \node[gnode] (j2) at (1.5, 1) {};
        \node[gnode] (j3) at (2.5, 1) {};

        \node[gnode] (k1) at (0, 0.5) {};
        \node[gnode] (l1) at (1, 0.5) {};
        \node[gnode] (m1) at (2, 0.5) {};
        \node[gnode] (k2) at (0, 1.5) {};
        \node[gnode] (l2) at (1, 1.5) {};
        \node[gnode] (m2) at (2, 1.5) {};
      \end{scope}

      \draw[q->] (i1) -- (l1);
      \draw (i1) -- ++(-135:{0.25*sqrt(2)} );
      \draw[q<-] (i1) -- ++(-45:{0.25*sqrt(2)} );
      \draw[q->] (i2) -- (m1);
      \draw (i2) -- ++(-135:{0.25*sqrt(2)} );
      \draw[q<-] (i2) -- ++(-45:{0.25*sqrt(2)} );
      \draw (i3) -- ++(-135:{0.25*sqrt(2)} );
      \draw[q<-] (i3) -- ++(-45:{0.25*sqrt(2)} );
      \draw (i3) -- ++(45:{0.25*sqrt(2)} );
      \draw[q->] (k1) -- (i1);
      \draw[q<-] (k1) -- ++(-135:{0.25*sqrt(2)} );
      \draw (k1) -- ++(135:{0.25*sqrt(2)} );
      \draw[q->] (l1) -- (i2);
      \draw[q->] (l1) -- (j1);
      \draw[q->] (m1) -- (i3);
      \draw[q->] (m1) -- (j2);
      \draw[q->] (j1) -- (k1);
      \draw[q->] (j1) -- (l2);
      \draw[q->] (j2) -- (m2);
      \draw[q->] (j2) -- (l1);
      \draw[q->] (j3) -- (m1);
      \draw (j3) -- ++(45:{0.25*sqrt(2)} );
      \draw[q<-] (j3) -- ++(-45:{0.25*sqrt(2)} );
      \draw[q->] (k2) -- (j1);
      \draw[q<-] (k2) -- ++(45:{0.25*sqrt(2)} );
      \draw (k2) -- ++(135:{0.25*sqrt(2)} );
      \draw[q<-] (k2) -- ++(-135:{0.25*sqrt(2)} );
      \draw[q->] (l2) -- (j2);
      \draw (m2) -- ++(135:{0.25*sqrt(2)} );
      \draw[q<-] (l2) -- ++(45:{0.25*sqrt(2)} );
      \draw (l2) -- ++(135:{0.25*sqrt(2)} );
      \draw[q->] (m2) -- (j3);
      \draw[q<-] (m2) -- ++(45:{0.25*sqrt(2)} );

      \draw[frame] (0,0) rectangle (3, 2);
    \end{tikzpicture}
  }
  \subfloat[\label{fig:BT-T}]{
    \begin{tikzpicture}[rotate=180]
      \fill[ws] (0,0) rectangle (3, 2);

      \fill[lshaded] (0,0.3) -- (0.2,0.3) -- (0.2,0.7) -- (0,0.5) -- cycle;
      \fill[lshaded] (0,1.3) -- (0.2,1.3) -- (0.2,1.7) -- (0,1.5) -- cycle;
      \fill[lshaded] (0.8,0.3) -- (1.2,0.3) -- (1.2,0.7) -- cycle;
      \fill[lshaded] (0.8,1.3) -- (1.2,1.3) -- (1.2,1.7) -- cycle;
      \fill[lshaded] (1.8,0.3) -- (2.2,0.3) -- (2.2,0.7) -- cycle;
      \fill[lshaded] (1.8,1.3) -- (2.2,1.3) -- (2.2,1.7) -- cycle;
      \fill[lshaded] (2.8,0.3) -- (3,0.3) -- (3,0.5) -- cycle;
      \fill[lshaded] (2.8,1.3) -- (3,1.3) -- (3,1.5) -- cycle;

      \fill[dshaded] (0.2,0) -- (0.5,0) -- (0.8,0.3) -- (0.2,0.3) -- cycle;
      \fill[dshaded] (1.2,0) -- (1.5,0) -- (1.8,0.3) -- (1.2,0.3) -- cycle;
      \fill[dshaded] (2.2,0) -- (2.5,0) -- (2.8,0.3) -- (2.2,0.3) -- cycle;
      \fill[dshaded] (0.2,0.7) -- (0.8,1.3) -- (0.2,1.3) -- cycle;
      \fill[dshaded] (1.2,0.7) -- (1.8,1.3) -- (1.2,1.3) -- cycle;
      \fill[dshaded] (2.2,0.7) -- (2.8,1.3) -- (2.2,1.3) -- cycle;
      \fill[dshaded] (0.2,1.7) -- (0.5,2) -- (0.2,2) -- cycle;
      \fill[dshaded] (1.2,1.7) -- (1.5,2) -- (1.2,2) -- cycle;
      \fill[dshaded] (2.2,1.7) -- (2.5,2) -- (2.2,2) -- cycle;

      \begin{scope}[shift={(0.2, 0)}]
        \draw[z-] (0, 0) -- (0, 2);
        \draw[z-] (1, 0) -- (1, 2);
        \draw[z-] (2, 0) -- (2, 2);
      \end{scope}
      
      \begin{scope}[shift={(0, 0.3)}]
        \draw[z-] (0, 0) -- (3, 0);
        \draw[z-] (0, 1) -- (3, 1);
      \end{scope}
      
      \draw[z-] (0.5, 0) -- (2.5, 2);
      \draw[z-] (1.5, 0) -- (3, 1.5);
      \draw[z-] (2.5, 0) -- (3, 0.5);
      \draw[z-] (0, 0.5) -- (1.5, 2);
      \draw[z-] (0, 1.5) -- (0.5, 2);

      \draw[boundary] (0,0) -- (3,0);
      \draw[boundary] (0,2) -- (3,2);
      \draw[frame] (0,0) -- (0,2);
      \draw[frame] (3,0) -- (3,2);
    \end{tikzpicture}
  }
  \subfloat[\label{fig:BT-T-quiver}]{
    \begin{tikzpicture}[rotate=180]
      \begin{scope}[shift={(0,0)}]
        \clip (0,0) rectangle (3,2);

        \node[fnode] (a) at (0.5,0.2) {};
        \node[fnode] (b) at (1.5,0.2) {};
        \node[fnode] (c) at (2.5,0.2) {};
        \node[gnode] (d) at (0.5,1) {};
        \node[gnode] (e) at (1.5,1) {};
        \node[gnode] (f) at (2.5,1) {};
        \node[fnode] (g) at (0.5,1.8) {};
        \node[fnode] (h) at (1.5,1.8) {};
        \node[fnode] (i) at (2.5,1.8) {};

        \node[fnode] (a') at (-0.5,0.2) {};
        \node[fnode] (d') at (-0.5,1) {};
        \node[fnode] (g') at (-0.5,1.8) {};
        \node[fnode] (f') at (3.5,1) {};
        \node[fnode] (i') at (3.5,1.8) {};
        
        \draw[q<-] (d) -- (a');
        \draw[q<-] (f') -- (c);
        \draw[q<-] (d') -- (d);
        \draw[q<-] (g) -- (d');
        \draw[q<-] (f) -- (f');
        \draw[q<-] (i') -- (f);
        \draw[q<-] (g') -- (g);
        \draw[q<-] (i) -- (i');
        
        \draw[q<-] (a) -- (d);
        \draw[q<-] (d) -- (e);
        \draw[q<-] (e) -- (a);

        \draw[q<-] (b) -- (e);
        \draw[q<-] (e) -- (f);
        \draw[q<-] (f) -- (b);
      
        \draw[q<-] (c) -- (f);

        \draw[q<-] (d) -- (g);
        \draw[q<-] (h) -- (d);

        \draw[q<-] (e) -- (h);
        \draw[q<-] (i) -- (e);
      
        \draw[q<-] (f) -- (i);

        \draw[q<-] (g) -- (h);
        \draw[q<-] (h) -- (i);
      \end{scope}

      \draw[boundary] (0,0) -- (3,0);
      \draw[boundary] (0,2) -- (3,2);
      \draw[frame] (0,0) -- (0,2);
      \draw[frame] (3,0) -- (3,2);
    \end{tikzpicture}
  }
  \caption{(a) A brane tiling on a torus.  (b) The periodic quiver
    associated with (a).  (c) A brane tiling on a finite-length
    cylinder.  The horizontal direction is periodic.  (d) The quiver
    for (c).}
  \label{fig:BT}
\end{figure}

As we just explained, a brane tiling gives rise to a four-dimensional
$\CN = 1$ theory.  A concrete description of this theory is known for
the subset of brane tilings that involve only $(N,0)$ 5-branes (i.e.\
$N$ coincident D5-branes) and $(N,\pm1)$ 5-branes.\cite{Franco:2005rj}
Given a brane tiling in this subset, we indicate $(N,1)$ and $(N,-1)$
5-brane regions by dark and light shading, respectively, while leaving
$(N,0)$ regions unshaded.  After the shading, we get a
checkerboard-like pattern on $\Sigma$ where shaded faces adjoin
unshaded ones and two shaded faces sharing a vertex are of different
types.%
\footnote{It is more common to represent such a brane tiling by a
  bipartite graph, placing a white node in each $(N,1)$ region and a
  black node in each $(N,-1)$ region, and connecting every pair of
  black and white nodes placed in two shaded regions sharing a vertex.
  The term ``zigzag paths'' originated in this context; we can draw
  them as lines running zigzag to avoid these nodes.}
Examples are shown in Fig.~\ref{fig:BT}.

Each unshaded region supports $N$ D5-branes, hence an $\SU(N)$ vector
multiplet lives there.  If the region contains part of the boundary,
the multiplet is frozen by boundary conditions and the associated
symmetry is an $\SU(N)$ flavor symmetry; otherwise it is dynamical and
we have an $\SU(N)$ gauge symmetry.  Adopting the quiver notation, we
represent a dynamical vector multiplet by a gauge node~$\gnode{}$ and
a nondynamical one by a flavor node~$\fnode{}$\,.

From open strings stretched between two unshaded regions sharing a
vertex, we get a chiral multiplet that transforms in the fundamental
representation under one of the associated gauge or flavor groups and
in the antifundamental representation under the other.  We represent
it by an arrow between the two nodes:
\begin{equation}
  \label{eq:arrow}
  \begin{tikzpicture}[scale={1/sqrt(2)}, baseline=(x.base)]
    \node (x) at (0,0) {\vphantom{x}};

    \fill[ws] (-135:1) -- (-45:1) -- (45:1) -- (135:1) -- cycle;

    \fill[dshaded] (0,0) -- (45:1) -- (-45:1) -- cycle;
    \fill[lshaded] (0,0) -- (135:1) -- (-135:1) -- cycle;
    
    \draw[z-] (-135:1) node[below] {$i$} -- (45:1);
    \draw[z-] (-45:1) node[below] {$j$} -- (135:1);
  \end{tikzpicture}
  \leadsto \
  \begin{tikzpicture}[baseline=(x.base)]
    \node (x) at (0,0) {\vphantom{x}};

    \node[fnode, above] (z) at (0,-0.5) {};
    \node[fnode, below] (w) at (0,0.5) {};
    \draw[q->] (z) -- (w);
  \end{tikzpicture}
  \ .
\end{equation}
The arrow points from the antifundamental side to the fundamental
side.  See Fig.~\ref{fig:BT} for examples of quivers obtained from
brane tilings.

Moreover, for every set of zigzag paths bounding a shaded region, we
have a loop of arrows and worldsheet instantons generate a
superpotential term given by the trace of the product of the
bifundamental chiral multiplets in the loop.  The coefficient of this
term is positive or negative depending on whether the direction of the
loop is clockwise or counterclockwise.  Thus, the four-dimensional
theory realized by a brane tiling in the subset under consideration is
an $\CN=1$ supersymmetric gauge theory described by a quiver with
potential drawn on $\Sigma$.

Each $\NS5_i$ supports a $\U(1)$ flavor symmetry $\U(1)_i$.  An arrow
is charged under $\U(1)_i$ if it is crossed by $C_i$.  The charge
$F_i$ of $\U(1)_i$ can be normalized in such a way that the arrow
in~\eqref{eq:arrow} has $F_i = -1$ and $F_j = +1$.  The diagonal
combination of all $\U(1)_i$ acts on the theory trivially since every
arrow is crossed by exactly two zigzag paths from the opposite sides.

The theory also has an R-symmetry $\U(1)_R$.  Its definition is not
unique as the R-charge $R$ can be shifted by a linear combination of
$\U(1)$ flavor charges.  However, the R-charge assignment is
constrained by two conditions.  The first is that $\U(1)_R$ must be
unbroken by the superpotential and therefore the R-charges of the
chiral multiplets contained in each superpotential term must add up to
two.  The second is that $\U(1)_R$ must be free of anomaly.  This
requires that for every gauge node, the sum of the R-charges of the
arrows starting from or ending at that node must equal the number of
the arrows minus two.

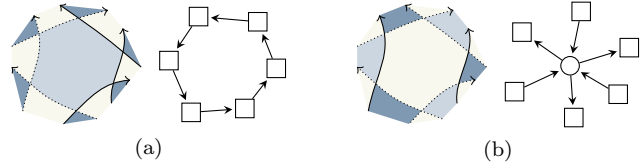
\begin{figure}
  \centering 
  \def\pathA{(0.3,0) .. controls (-1,0.7) .. (-2,2)}
  \def\pathB{(-1.7,0.3) .. controls (-1.5,1.8) and (-1,2.5) .. (-1.4,4)}
  \def\pathC{(-2,2.8) .. controls (-0.5,3.8) .. (0.6,4.2)}
  \def\pathD{(-0.7,4.4) .. controls (0.5,3.2) .. (1.5,2)}
  \def\pathE{(1,3.5) .. controls (1.2,2.5) and (0.6,1.5) .. (0.7,0.6)}
  \def\pathF{(1.2,1.5) .. controls (0.5,1.4) and (0.2,1) .. (-0.6,-0.2)}

  \subfloat[\label{fig:zigzag-loop-a}]{
    \begin{tikzpicture}[xscale=0.5, yscale=0.35]
      \path[name path=A] \pathA;
      \path[name path=B] \pathB;
      \path[name path=C] \pathC;
      \path[name path=D] \pathD;
      \path[name path=E] \pathE;
      \path[name path=F] \pathF;

      \begin{scope}
        \clip \pathA -- \pathC -- \pathE -- cycle;
        \fill[ws] (-2,-0.5) rectangle (2,5);
      \end{scope}

      \begin{scope}
        \clip \pathB -- \pathD -- \pathF -- cycle;
        \fill[ws] (-2,-0.5) rectangle (2,5);
      \end{scope}

      \begin{scope}
        \clip
        \pathA -- \pathB -- \pathC -- \pathD -- \pathE -- \pathF -- cycle;

        \fill[dshaded] (-2,-0.5) rectangle (2,5);
      \end{scope}

      \path[name intersections={of=A and B, by=1}];
      \path[name intersections={of=B and C, by=2}];
      \path[name intersections={of=C and D, by=3}];
      \path[name intersections={of=D and E, by=4}];
      \path[name intersections={of=E and F, by=5}];
      \path[name intersections={of=F and A, by=6}];

      \begin{scope}
        \clip \pathA -- \pathB -- \pathC -- \pathD -- \pathE -- \pathF -- cycle;

        \clip
        (1) -- ++(95:0.5) --
        (2) -- ++(75:0.5) --
        (3) -- ++(-30:0.5) --
        (4) -- ++(-90:0.5) --
        (5) --
        (6) -- ++(180:0.5) -- cycle;

        \fill[lshaded] (-2,-0.5) rectangle (2,5);
      \end{scope}

      \draw[wz->] \pathA;
      \draw[wz->] \pathB;
      \draw[wz->] \pathC;
      \draw[z<-] \pathD;
      \draw[z<-] \pathE;
      \draw[z<-] \pathF;
    \end{tikzpicture}
    \
    \begin{tikzpicture}[xscale=0.5, yscale=0.35]
      \node[fnode] (a) at (0.35,0.45) {};
      \node[fnode] (b) at (1.2,2) {};
      \node[fnode] (c) at (0.7,3.8) {};
      \node[fnode] (d) at (-0.9,4) {};
      \node[fnode] (e) at (-1.7,2.3) {};
      \node[fnode] (f) at (-1.1,0.3) {};
      \draw[q->] (a) -- (b);
      \draw[q->] (b) -- (c);
      \draw[q->] (c) -- (d);
      \draw[q->] (d) -- (e);
      \draw[q->] (e) -- (f);
      \draw[q->] (f) -- (a);
    \end{tikzpicture}
  }
  \qquad
  \subfloat[\label{fig:zigzag-loop-b}]{
    \begin{tikzpicture}[xscale=0.5, yscale=0.35]
      \path[name path=A] \pathA;
      \path[name path=B] \pathB;
      \path[name path=C] \pathC;
      \path[name path=D] \pathD;
      \path[name path=E] \pathE;
      \path[name path=F] \pathF;

      \begin{scope}
        \clip \pathA -- \pathC -- \pathE -- cycle;
        \fill[lshaded] (-2,-0.5) rectangle (2,5);
      \end{scope}

      \begin{scope}
        \clip \pathB -- \pathD -- \pathF -- cycle;
        \fill[dshaded] (-2,-0.5) rectangle (2,5);
      \end{scope}

      \begin{scope}
        \clip \pathA -- \pathB -- \pathC -- \pathD -- \pathE -- \pathF -- cycle;
        \fill[ws] (-2,-0.5) rectangle (2,5);
      \end{scope}

      \draw[wz->] \pathA;
      \draw[z->] \pathB;
      \draw[wz->] \pathC;
      \draw[wz<-] \pathD;
      \draw[z<-] \pathE;
      \draw[wz<-] \pathF;
    \end{tikzpicture}
    \
    \begin{tikzpicture}[xscale=0.5, yscale=0.35]
      \node[gnode] (g) at (-0.3,2.1) {};
      \node[fnode] (a) at (-0.2,0.1) {};
      \node[fnode] (b) at (1,1) {};
      \node[fnode] (c) at (1.3,2.8) {};
      \node[fnode] (d) at (0,4.2) {};
      \node[fnode] (e) at (-1.6,3.4) {};
      \node[fnode] (f) at (-1.8,1.1) {};
      
      \draw[q->] (g) -- (a);
      \draw[q<-] (g) -- (b);
      \draw[q->] (g) -- (c);
      \draw[q<-] (g) -- (d);
      \draw[q->] (g) -- (e);
      \draw[q<-] (g) -- (f);
    \end{tikzpicture}
  }
  \caption{Zigzag paths bounding (a) a shaded region and (b) an
    unshaded region.  In either case, the R-charges of two of the
    arrows are different from those of the rest.}
  \label{fig:zigzag-loop}
\end{figure}

To fix the R-charge assignment, let us assume that we can orient the
zigzag paths (and deform them if necessary) and bound every shaded or
unshaded region with zigzag paths all heading upward, for some choice
of the ``vertical'' direction in the neighborhood of that region.
This is the case for the examples in Fig.~\ref{fig:BT}.  The zigzag
paths thus oriented fall into two groups; when a zigzag path goes
upward and we cross it from the left to the right, $q$ increases or
decreases by one.  We distinguish the latter case from the former by
drawing the zigzag path with a dotted line.  Then, we give an arrow
$R = 0$ if it originates from a crossing of two zigzag paths of the
same type, and $R=1$ otherwise.  With this R-charge assignment the two
conditions described above are satisfied; see
Fig.~\ref{fig:zigzag-loop} for illustration.

\begin{figure}[b]
  \centering
  \subfloat[\label{fig:quiver-rule-a}]{
    \begin{tikzpicture}[scale={1/sqrt(2)}, baseline=(x.base)]
      \node (x) at (0,0) {\vphantom{x}};

      \fill[ws] (-135:1) -- (-45:1) -- (45:1) -- (135:1) -- cycle;

      \fill[dshaded] (0,0) -- (45:1) -- (-45:1) -- cycle;
      \fill[lshaded] (0,0) -- (135:1) -- (-135:1) -- cycle;

      \draw[z->] (-135:1) node[below] {$i$} -- (45:1);
      \draw[z->] (-45:1) node[below] {$j$} -- (135:1);
    \end{tikzpicture}
    \
    \begin{tikzpicture}[baseline=(x.base)]
      \node (x) at (0,0) {\vphantom{x}};

      \node[fnode, above] (z) at (0,-0.5) {};
      \node[fnode, below] (w) at (0,0.5) {};
      \draw[q->] (z) -- (w);
    \end{tikzpicture}
  }
  \quad
  \subfloat[\label{fig:quiver-rule-b}]{
    \begin{tikzpicture}[scale={1/sqrt(2)}, baseline=(x.base)]
      \node (x) at (0,0) {\vphantom{x}};

      \fill[ws] (-135:1) -- (-45:1) -- (45:1) -- (135:1) -- cycle;

      \fill[lshaded] (0,0) -- (45:1) -- (-45:1) -- cycle;
      \fill[dshaded] (0,0) -- (135:1) -- (-135:1) -- cycle;

      \draw[wz->] (-135:1) node[below] {$i$} -- (45:1);
      \draw[wz->] (-45:1) node[below] {$j$} -- (135:1);
    \end{tikzpicture}
    \
    \begin{tikzpicture}[baseline=(x.base)]
      \node (x) at (0,0) {\vphantom{x}};

      \node[fnode, above] (z) at (0,-0.5) {};
      \node[fnode, below] (w) at (0,0.5) {};
      \draw[q<-] (z) -- (w);
    \end{tikzpicture}
  }
  \quad
  \subfloat[\label{fig:quiver-rule-c}]{
    \begin{tikzpicture}[scale={1/sqrt(2)}, baseline=(x.base)]
      \node (x) at (0,0) {\vphantom{x}};

      \fill[ws] (-135:1) -- (-45:1) -- (45:1) -- (135:1) -- cycle;
    
      \fill[lshaded] (45:1) -- (0,0) -- (135:1) -- cycle;
      \fill[dshaded] (-45:1) -- (0,0) -- (-135:1) -- cycle;

      \draw[z->] (-135:1) node[below] {$i$} -- (45:1);
      \draw[wz->] (-45:1) node[below] {$j$} -- (135:1);
    \end{tikzpicture}
    \begin{tikzpicture}[baseline=(x.base)]
      \node (x) at (0,0) {\vphantom{x}};

      \node[fnode, right] (z) at (-0.5,0) {};
      \node[fnode, left] (w) at (0.5,0) {};
      \draw[q->] (z) -- (w);
    \end{tikzpicture}
  }
  \quad
  \subfloat[\label{fig:quiver-rule-d}]{
    \begin{tikzpicture}[scale={1/sqrt(2)}, baseline=(x.base)]
      \node (x) at (0,0) {\vphantom{x}};

      \fill[ws] (-135:1) -- (-45:1) -- (45:1) -- (135:1) -- cycle;

      \fill[dshaded] (45:1) -- (0,0) -- (135:1) -- cycle;
      \fill[lshaded] (-45:1) -- (0,0) -- (-135:1) -- cycle;

      \draw[wz->] (-135:1) node[below] {$i$} -- (45:1);
      \draw[z->] (-45:1) node[below] {$j$} -- (135:1);
    \end{tikzpicture}
    \begin{tikzpicture}[baseline=(x.base)]
      \node (x) at (0,0) {\vphantom{x}};

      \node[fnode, right] (z) at (-0.5,0) {};
      \node[fnode, left] (w) at (0.5,0) {};
      \draw[q<-] (z) -- (w);
    \end{tikzpicture}
  }
  \caption{The rule for assigning a quiver to a brane tiling diagram.
    The arrows in (a) and (b) have $(R, F_i, F_j) = (0, -1, 1)$.
    Those in (c) and (d) have $(R, F_i, F_j) = (1, 1, -1)$.}
  \label{fig:quiver-rule}
\end{figure}
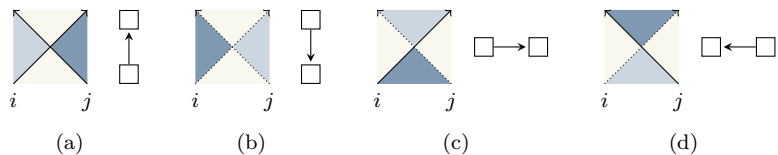

The rule for reading off the quiver from zigzag paths is summarized
in~Fig.~\ref{fig:quiver-rule}.

\subsection{Integrable lattice models arising from brane tilings}

From the supersymmetric index of the four-dimensional $\CN = 1$ theory
realized by a brane tiling, we obtain an integrable lattice model
defined on the lattice $\{C_i\}$ consisting of the zigzag paths.  Each
$C_i$ carries a spectral parameter $u_i$.  S-duality followed by
T-duality on $S^1$ turns $\NS5_i$ into a D4-brane, and its coordinate
on the dual circle $\St^1$ is $u_i$.  Instead, we can apply T-duality
on $S^1$ and lift $\NS5_i$ to an M5-brane; then $u_i$ is the
coordinate on the M-theory circle.  Either way, $u_i$ is determined by
the holonomy of the $\U(1)$ gauge field on $\NS5_i$ around $S^1$.

If the theory is described by a quiver,
translation between the gauge theory and the lattice model goes as
follows.\cite{Spiridonov:2010em}

Nodes are interpreted as spin sites.  For each flavor node, we can
turn on a holonomy of the associated gauge field.  The index depends
on the conjugacy class of the holonomy, which is uniquely represented
by a diagonal matrix $\diag(z_1, \dotsc, z_N)$ up to permutations of
the entries.  The index is therefore a symmetric function of the
$\U(1)$-valued variables $(z_1, \dotsc, z_N)$ obeying the constraint
$z_1 \dotsm z_N = 1$.  These variables are fugacities for the $\SU(N)$
flavor symmetry and parameterize the value of the spin at this node;
thus spins take values in the maximal torus $\U(1)^{N-1}$ of $\SU(N)$.
For a gauge node, integration is performed over the fugacities since
its gauge field is a path integral variable.  This is the summation
over the values of a spin placed on an internal face.  Finally, arrows
represent interactions between spins.

The unitarity relations are satisfied if the contributions to the
index from arrows with $R=0$ are properly normalized.  For example,
consider the relation%
\footnote{By an equality of two quivers, we mean that the
  supersymmetric indices of the theories described by those quivers
  are equal.}
\begin{equation}
  \label{eq:unitarity1}
  \def\pathA{(0,0) to[out=0, in=180] (1,0.5) to[out=0, in=180] (2,0)}
  \def\pathB{(2,0.5) to[in=0,out=180] (1,0) to[in=0,out=180] (0,0.5)}
  \begin{tikzpicture}[scale=0.9, baseline=(x.base)]
    \node (x) at (1,0.25) {\vphantom{x}};

    \begin{scope}
      \clip \pathA -- (2,-0.25) -- (0,-0.25) -- cycle;
      \fill[dshaded] (0,-0.25) rectangle (2,0.25);
    \end{scope}

    \begin{scope}
      \clip \pathB -- (0,0.75) -- (2,0.75) -- cycle;
      \fill[lshaded] (0,0.25) rectangle (2,0.75);
    \end{scope}

    \begin{scope}
      \clip \pathA -- \pathB -- cycle;
      \fill[ws] (0,0) rectangle (2,0.5);
    \end{scope}

    \draw[z->] \pathA;
    \draw[z<-] \pathB;
  \end{tikzpicture}
  \ = \
  \begin{tikzpicture}[scale=0.9, baseline=(x.base)]
    \node (x) at (1,0.25) {\vphantom{x}};

    \fill[ws] (0,0) rectangle (2,0.5);

    \fill[dshaded] (0,-0.25) rectangle (2,0);
    \fill[lshaded] (0,0.5) rectangle (2,0.75);

    \draw[z->] (0,0) -- (2,0);
    \draw[z->] (0,0.5) -- (2,0.5);
  \end{tikzpicture}
  \iff
  \begin{tikzpicture}[scale=0.9, baseline=(x.base)]
    \node (x) at (1,0.25) {\vphantom{x}};

    \node[fnode, right] (1) at (0,0.25) {};
    \node[gnode] (2) at (1,0.25) {};
    \node[fnode, left] (3) at (2,0.25) {};
    \draw[q->] (1) -- (2);
    \draw[q->] (2) -- (3);
  \end{tikzpicture}
  \ = \
  \begin{tikzpicture}[scale=0.9, baseline=(x.base)]
    \node (x) at (0.5,0.25) {\vphantom{x}};

    \node[fnode] (1) at (0,0.25) {};
    \node[fnode] (2) at (1,0.25) {};
    \draw[eq-] (1) -- (2);
  \end{tikzpicture}
  \ ,
\end{equation}
where the right-hand side is a ``delta function'' that equates two
flavor nodes when one of them is gauged.  The theory on the left-hand
side is SQCD with $N$ colors and $N$ flavors.  It exhibits confinement
and has a vacuum in which the mesons take nonzero expectation values
and the flavor symmetry $\SU(N) \times \SU(N)$ is broken to the
diagonal subgroup.~\cite{Seiberg:1994bz} The index computed in this
vacuum is given by the right-hand side, provided that we cancel the
contributions from the surviving baryon and antibaryon.  Another
relation
\begin{equation}
  \label{eq:BB=0}
  \def\pathA{(0,0) to[out=0, in=180] (1,0.5) to[out=0, in=180] (2,0)}
  \def\pathB{(2,0.5) to[in=0,out=180] (1,0) to[in=0,out=180] (0,0.5)}
  \begin{tikzpicture}[scale=0.9, baseline=(x.base)]
    \node (x) at (1,0.25) {\vphantom{x}};

    \fill[ws] (0,-0.25) rectangle (2,0.75);

    \path[name path=A] \pathA;
    \path[name path=B] \pathB;
    \path[name intersections={of=A and B, by={1,2}}];

    \begin{scope}
      \clip \pathA -- \pathB -- cycle;
      \fill[dshaded] (0,0) -- (0.5,0) -- (1) -- (0.5,0.5) -- (0,0.5) 
      -- cycle;
    \end{scope}

    \begin{scope}
      \clip \pathA -- \pathB -- cycle;
      \fill[dshaded]  (2,0) -- (1.5,0) -- (2) -- (1.5,0.5) -- (2,0.5) 
      -- cycle;
     \end{scope}

    \begin{scope}
      \clip \pathA -- \pathB -- cycle;
      \fill[lshaded] (0.5,0) -- (1.5,0) -- (2) -- (1.5,0.5) -- (0.5,0.5)
      -- (1) -- (0.5,0) -- cycle;
     \end{scope}

    \draw[wz->] \pathA;
    \draw[z<-] \pathB;
  \end{tikzpicture}
  \ = \
  \begin{tikzpicture}[scale=0.9, baseline=(x.base)]
    \node (x) at (1,0.25) {\vphantom{x}};

    \fill[ws] (0,-0.25) rectangle (2,0.75);

    \fill[dshaded] (0,0) rectangle (2,0.5);

    \draw[wz->] (0,0) -- (2,0);
    \draw[z->] (0,0.5) -- (2,0.5);
  \end{tikzpicture}
  \iff
  \begin{tikzpicture}[scale=0.9, baseline=(x.base)]
    \node (x) at (0,0.25) {\vphantom{x}};

    \node[fnode] (1) at (0,0.7) {};
    \node[fnode] (2) at (0,-0.2) {};
    \draw[q->] (1) to[bend left] (2);
    \draw[q->] (2) to[bend left] (1);
  \end{tikzpicture}
  \ = \ 
  \begin{tikzpicture}[scale=0.9, baseline=(x.base)]
    \node (x) at (0,0.25) {\vphantom{x}};

    \node[fnode] (1) at (0,0.7) {};
    \node[fnode] (2) at (0,-0.2) {};
  \end{tikzpicture}
\end{equation}
holds since the two arrows on the left-hand side form a loop and
generates a mass term in the superpotential.  We can send the mass to
infinity so that these arrows decouple from the theory, and are left
with the right-hand side.

The Yang--Baxter equation with three zigzag paths is harder to
understand, as it always involves an $(N,q)$ region with $|q| > 1$ and
a quiver description is not available.  The problem stems from the
fact that our defects are domain walls across which $q$ changes.  To
circumvent the difficulty, we take a pair of zigzag paths of different
types and think of it as a single line:
\begin{equation}
  \label{eq:solid-line}
  \begin{tikzpicture}[baseline=(x.base)]
    \node (x) at (0.5,0) {\vphantom{x}};

    \draw[r->] (0,0)-- (1,0);
  \end{tikzpicture}
  \ = \
  \begin{tikzpicture}[baseline=(x.base)]
    \node (x) at (0.5,0) {\vphantom{x}};

    \draw[z->] (0,0.125) -- (1,0.125);
    \draw[wz->] (0,-0.125) -- (1,-0.125);
  \end{tikzpicture}
  \ .
\end{equation}
This line does not alter the value of $q$.  Taking two copies of this
line and placing them in an $(N,-1)$ background, we can make the
R-matrix
\begin{equation}
  \label{eq:RD}
  \begin{tikzpicture}[baseline=(x.base), scale=0.75]
    \node (x) at (1,1) {\vphantom{x}};

    \fill[lshaded] (0,0) rectangle (2,2);
    \draw[r->] (0,1) -- (2,1);
    \draw[r->] (1,0) -- (1,2);
  \end{tikzpicture}
  \ = \
  \begin{tikzpicture}[baseline=(x.base), scale=0.75]
    \node (x) at (1,1) {\vphantom{x}};
    
    \fill[ws] (0,0) rectangle (2,2);

    \fill[lshaded] (0,0) rectangle (0.833,0.833);
    \fill[lshaded] (1.166,0) rectangle (2,0.833);
    \fill[lshaded] (0,1.166) rectangle (0.833,2);
    \fill[lshaded] (1.166,1.166) rectangle (2,2);
    \fill[dshaded] (0.833,0.833) rectangle (1.166,1.166);

    \draw[z->] (0,1.166) -- (2,1.166);
    \draw[wz->] (0,0.833) -- (2,0.833);
    \draw[z->] (0.833,0) -- (0.833,2);
    \draw[wz->] (1.166,0) -- (1.166,2);
  \end{tikzpicture}
  \ = \
  \begin{tikzpicture}[baseline=(x.base), scale=0.75]
    \node (x) at (1,1) {\vphantom{x}};

    \node[fnode, above] (zj) at (1,0) {};
    \node[fnode, right] (zi) at (0,1) {};
    \node[fnode, below] (wj) at (1,2) {};
    \node[fnode, left] (wi) at (2,1) {};

    \draw[q->] (zj) -- (zi);
    \draw[q->] (zi) -- (wj);
    \draw[q->] (wj) -- (wi);
    \draw[q->] (wi) -- (zj);
  \end{tikzpicture}
  \ .
\end{equation}
A lattice model constructed from this R-matrix is a vertex model whose
quiver consists of diamonds of arrows; see Fig.~\ref{fig:BT-D-quiver}.
The vector space carried by a line is the space of symmetric functions
of fugacities $(z_1, \dotsc, z_N)$.

Alternatively, we can place these lines in an $(N,0)$ background and
force them to exchange their constituent zigzag paths as they cross:
\begin{equation}
  \label{eq:RT}
    \begin{tikzpicture}[baseline=(x.base), scale=0.75]
      \node (x) at (1,1) {\vphantom{x}};
      
      \fill[ws] (0,0) rectangle (2,2);

      \draw[r->] (0,1) -- (2,1);
      \draw[r->] (1,0) -- (1,2);
    \end{tikzpicture}
    \ = \
    \begin{tikzpicture}[baseline=(x.base), scale=0.75]
      \node (x) at (1,1) {\vphantom{x}};
      
      \fill[ws] (0,0) rectangle (2,2);

      \fill[lshaded] (1,0.666) rectangle (1.333,1);
      \fill[dshaded] (0.666,0) rectangle (1,0.666);
      \fill[dshaded] (1.333,1) rectangle (2,1.333);
      \fill[dshaded] (0,1) -- (1,1) -- (1,2) -- (0.666,2) -- (0.666,1.333)
      -- (0,1.333) -- cycle;
      
      \draw[wz->] (1,0) -- (1,2);
      \draw[wz->] (0,1) -- (2,1);
      
      \draw[z->] (0.666,0) -- (0.666,0.666) -- (1.333,0.666) -- (1.333,1.333) -- (2,1.333);
      \draw[z->] (0,1.333) -- (0.666,1.333) -- (0.666,2);
    \end{tikzpicture}
    \ = \
    \begin{tikzpicture}[baseline=(x.base), scale=0.75]
      \node (x) at (1,1) {\vphantom{x}};
      
      \node[fnode] (zj) at (0.35,0.35) {};
      \node[fnode] (wi) at (1.65,0.35) {};
      \node[fnode] (wj) at (1.65,1.65) {};
      
      \draw[q->] (zj) -- (wi);
      \draw[q->] (wi) -- (wj);
      \draw[q->] (wj) -- (zj);
    \end{tikzpicture}
    \ .
\end{equation}
This R-matrix leads to an IRF model described by a quiver with
triangles of arrows, as shown in Fig.~\ref{fig:BT-T-quiver}.  The
corresponding Yang--Baxter equation, after cancellation of some
factors with the help of the unitarity relation \eqref{eq:BB=0}, reads
\begin{equation}
  \label{eq:Seiberg}
  \begin{tikzpicture}[baseline=(x.base)]
    \node (x) at (0,0) {\vphantom{x}};
    
    \node[fnode] (a) at (240:1){};
    \node[fnode] (c) at (0:1) {};
    \node[fnode] (d) at (60:1) {};
    \node[fnode] (f) at (180:1) {};
    \node[gnode] (g) at (0,0) {};
    
    \draw[q->] (a) -- (g);
    \draw[q->] (g) -- (c);
    \draw[q->] (d) -- (g);    
    \draw[q->] (g) -- (f);
  \end{tikzpicture}
  \ = \
  \begin{tikzpicture}[baseline=(x.base)]
    \node (x) at (0,0) {\vphantom{x}};
    
    \node[fnode] (a) at (240:1){};
    \node[fnode] (c) at (0:1) {};
    \node[fnode] (d) at (60:1) {};
    \node[fnode] (f) at (180:1) {};
    \node[gnode] (g) at (0,0) {};
    
    \draw[q->] (g) -- (a);
    \draw[q->] (c) -- (g);
    \draw[q->] (f) -- (g);
    \draw[q->] (g) -- (d);
    \draw[q->] (d) -- (f);
    
    \draw[q<-] (c) -- (a);
    \draw[q<-] (c) -- (d);
    \draw[q<-] (f) -- (a);
  \end{tikzpicture}
  \ .
\end{equation}
The two sides are related by Seiberg duality\cite{Seiberg:1994pq} for
SQCD with $N$ colors and $2N$ flavors, so their indices are indeed
equal.  The Yang--Baxter equation for the R-matrix~\eqref{eq:RD},
though more complicated, also follows from this equality.  The
relation between the Yang-Baxter move and Seiberg duality was pointed
out in Ref.~15.

\subsection{Surface defects as transfer matrices}

The brane tiling construction of integrable lattice models can be
enriched by introduction of surface defects.\cite{Maruyoshi:2016caf}
Consider a brane tiling configuration, and add to it a D3-brane that
creates a defect $\CE_{\D3}$ in $\thy_{\D5}$.  Let the support of
$\CE_{\D3}$ be $S^1 \times N \times C$, where $N$ is a curve in $M$
and $C$ is a closed curve on~$\Sigma$.  In the four-dimensional
$\CN=1$ theory, $\CE_{\D3}$ becomes a half-BPS surface defect
$\CE_{\D3}[C]$ on $S^1 \times N$.

In the lattice model, $\CE_{\D3}$ appears as a new oriented line,
which we represent by a dashed line.  Now that we have two kinds of
lines, we can define three R-matrices:
\begin{equation}
  \RM
  =
  \begin{tikzpicture}[scale=0.5, baseline=(x.base)]
    \node (x) at (1,1) {\vphantom{x}};
    \draw[r->] (0,1) -- (2,1);
    \draw[r->] (1,0) -- (1,2);
  \end{tikzpicture}
  \,,
  \qquad
  \LM
  =
  \begin{tikzpicture}[scale=0.5, baseline=(x.base)]
    \node (x) at (1,1) {\vphantom{x}};

    \draw[dr->] (0,1) -- (2,1);
    \draw[r->] (1,0) -- (1,2);
  \end{tikzpicture}
  \,,
  \qquad
  \CRM
  =
  \begin{tikzpicture}[scale=0.5, baseline=(x.base)]
    \node (x) at (1,1) {\vphantom{x}};

    \draw[dr->] (0,1) -- (2,1);
    \draw[dr->] (1,0) -- (1,2);
  \end{tikzpicture}
  \ .
\end{equation}
The middle one is called the \emph{L-operator}.  Correspondingly, we
have four Yang--Baxter equations, involving zero to three dashed
lines.  Those that mix different R-matrices,
\begin{equation}
  \label{eq:RLL}
  \begin{tikzpicture}[scale=0.5, baseline=(x.base)]
    \node (x) at (30:2) {\vphantom{x}};
    
    \draw[r->] (0,0) -- ++(30:3);
    \draw[dr->] (0,2) -- ++(-30:3);
    \draw[r->] (-30:1) -- ++(0,3);
  \end{tikzpicture}
  \ = \
  \begin{tikzpicture}[scale=0.5, baseline=(x.base)]
    \node (x) at (30:1) {\vphantom{x}};
    
    \draw[r->] (0,0) -- ++(30:3);
    \draw[dr->] (0,1) -- ++(-30:3);
    \draw[r->] (-30:2) -- ++(0,3);
  \end{tikzpicture}
  \quad
  \text{and}
  \quad
  \begin{tikzpicture}[scale=0.5, baseline=(x.base)]
    \node (x) at (30:2) {\vphantom{x}};
    
    \draw[dr->] (0,0) -- ++(30:3);
    \draw[dr->] (0,2) -- ++(-30:3);
    \draw[r->] (-30:1) -- ++(0,3);
  \end{tikzpicture}
  \ = \
  \begin{tikzpicture}[scale=0.5, baseline=(x.base)]
    \node (x) at (30:1) {\vphantom{x}};
    
    \draw[dr->] (0,0) -- ++(30:3);
    \draw[dr->] (0,1) -- ++(-30:3);
    \draw[r->] (-30:2) -- ++(0,3);
  \end{tikzpicture}
  \ ,
\end{equation}
are called \emph{RLL relations}.

The effect of the surface defect on the lattice model can be phrased
compactly in terms of the L-operator.  The neighborhood of the dashed
line looks like
\begin{equation}
  \label{eq:TM-L}
  \begin{tikzpicture}[scale=0.5]
    \node (x) at (1,0) {\vphantom{x}};

    \draw[dr->, right hook->] (-0.1,0) -- (1.1,0);
    \draw[dr->, >=left hook] (1,0) -- (4.1,0);
    \node[fill=white, inner sep=1pt] at (2.5,0) {$\,\dots$};

    \begin{scope}[shift={(0,-0.5)}]
      \draw[r->] (0.5,0) -- (0.5,1.2);
      \draw[r->] (1.5,0) -- (1.5,1.2);
      \draw[r->] (3.5,0) -- (3.5,1.2);
    \end{scope}
  \end{tikzpicture}
  \ .
\end{equation}
This picture shows that the surface defect acts on the lattice model
by a transfer matrix constructed from L-operators.

In fact, $\thy_{\D5}$ has a whole family of half-BPS defects.  Each of
them corresponds to an irreducible representation of $\SU(N)$ and is
constructed with D3-branes that stretch between the D5-branes and
extra NS5-branes.  The defect $\CE_{\D3}$ described above is the
simplest member of this family, corresponding to the fundamental
representation.  Thus, the four-dimensional theory has a family of
surface defects parametrized by the irreducible representations of
$\SU(N)$.  The insertion of a surface defect is mapped in the lattice
model to the action of a transfer matrix constructed from L-operators,
which contains a dashed line labeled with a representation in addition
to the spectral parameter and the curve~$N$.  The vector space carried
by the dashed line is the representation space.

\subsection{The three-sphere case}
\label{sec:m-=-s3}

To conclude our discussion, we describe the integrable lattice models
arising from brane tilings concretely for $M = S^3$, equipped with the
round metric of radius~$1$. Other cases are possible and interesting;
the case where $M$ is a lens space $L(r,1)$ was considered in Ref.~19.

Parametrize $S^3$ by two complex variables $(\zeta_p,\zeta_q)$
satisfying $|\zeta_p|^2 + |\zeta_q|^2 = 1$, and denote the isometry
groups acting on $\zeta_p$ and $\zeta_q$ by $\U(1)_p$ and $\U(1)_q$,
respectively.  We take $S^1 \times M$ to be a twisted product; we
prepare a trivial $S^3$-fibration over an interval $[0, \beta]$ and
identify the fibers at the ends of the base using an isometry
$(e^{i\theta_p}, e^{i\theta_q}) \in \U(1)_p \times \U(1)_q$.  On this
spacetime, the partition function of the quiver gauge theory realized
by a brane tiling gives the supersymmetric index refined by the
isometries and the flavor symmetries.\cite{Romelsberger:2005eg,
  Kinney:2005ej, Festuccia:2011ws}

The index can be computed exactly, for instance by localization of the
path integral.  The result is expressed in terms of the elliptic gamma
function
\begin{equation}
  \Gamma(z; p,q) 
  =
  \prod_{j,k=0}^\infty
  \frac{1 - z^{-1} p^{j+1} q^{k+1}}{1 - zp^j q^k}
\end{equation}
with $p = e^{-\beta + \iu\theta_p}$ and
$q = e^{-\beta + \iu\theta_q}$.\cite{Dolan:2008qi} To write down the
formula, let $a_i = e^{2\pi\iu u_i}$ be the fugacity for the flavor
group $\U(1)_i$ associated with the $i$th zigzag path.  Also, we
introduce the notation
$(z;q)_\infty = \prod_{k=0}^\infty (1 - q^k z)$.

A bifundamental chiral multiplet with R-charge $R$ and $\U(1)_i$
charge $F_i$ contributes to the index by the factor
\begin{equation}
  \prod_{I,J=1}^N
  \Gamma\biggl((pq)^{R/2} \prod_i a_i^{F_i} \frac{w_I}{z_J};
  p,q\biggr)
  \,,
\end{equation}
where $z_J$ are fugacities for the node at the tail of the arrow and
$w_I$ are those for the node at the tip.  To find the full index, we
take the product of the contributions from all arrows, and then for
each gauge node, integrate over its fugacities~$z_I$ with the measure
\begin{equation}
  \frac{(p;p)_\infty^{N-1} (q;q)_\infty^{N-1}}{N!}
  \prod_{I=1}^{N-1} \frac{\rmd z_I}{2\pi\iu z_I}
  \prod_{\substack{I, J = 1 \\ I \neq J}}^N
  \frac{1}{\Gamma(z_I/z_J; p,q)}
  \,.
\end{equation}
The integration contour is the unit circle for each fugacity.

The unitarity relation \eqref{eq:unitarity1} is satisfied if we
normalize the contribution from each arrow with $R=0$ by dividing it
by the factor $\Gamma(\prod_i a_i^{N F_i}; p,q)$, which cancels the
contribution from the corresponding baryon.  The Yang--Baxter
equation~\eqref{eq:Seiberg} is an integral identity\cite{MR2044635,
  MR2630038} obeyed by the elliptic gamma function.\cite{Dolan:2008qi}

There are two circles in $S^3$ around which we can place half-BPS
surface defects without breaking the isometries, namely
$\{\zeta_q = 0\}$ and $\{\zeta_p = 0\}$.  Accordingly, dashed lines
come in two types, related by an interchange of $p$ and $q$.  If a
dashed line is in an $n$-dimensional representation, the L-operator in
an $(N,-1)$ background
\begin{equation}
  \label{eq:Ldia-zz}
  \LM
  =
  \begin{tikzpicture}[scale=0.5, baseline=(x.base)]
    \node (x) at (1,1) {\vphantom{x}};

    \fill[lshaded] (0,0) rectangle (2,2);
    \draw[dr->] (0,1) -- (2,1);
    \draw[r->] (1,0) -- (1,2);
  \end{tikzpicture}
  =
  \begin{tikzpicture}[scale=0.5, baseline=(x.base)]
    \node (x) at (1,1) {\vphantom{x}};

    \fill[ws] (0,0) rectangle (2,2);

    \fill[lshaded] (0,0) rectangle (0.75,2);
    \fill[lshaded] (1.25,0) rectangle (2,2);
    \draw[dr->] (0,1) -- (2,1);
    \draw[z->] (0.75,0) -- (0.75,2);
    \draw[wz->]   (1.25,0) -- (1.25,2);
  \end{tikzpicture}
\end{equation}
may be represented as an $n \times n$ matrix, whose entries are
difference operators acting on the fugacities for the flavor node
associated with the $(N,0)$ region below the dashed line.  It
satisfies the RLL relation with the R-matrix~\eqref{eq:RD}.  This
R-matrix defines the Bazhanov--Sergeev model of type
$\SU(N)$~\cite{Bazhanov:2011mz}.

For the fundamental representation of $\SU(2)$, the above L-operator
was identified in Ref.~9.
It is essentially Sklyanin's L-operator,\cite{Sklyanin:1983ig} which
satisfies the RLL relation with Baxter's R-matrix for the eight-vertex
model.\cite{Baxter:1971cr, Baxter:1972hz}

For the fundamental representation of $\SU(N)$ with general $N$, we
get the L-operator for Belavin's elliptic R-matrix
\cite{Belavin:1981ix}.  If instead placed in an $(N,0)$ background,
the L-operator gives a representation of Felder's elliptic quantum
group for $\slf_N$.\cite{Felder:1994be, Felder:1994pb, MR1606760} The
details will be presented elsewhere.

\section*{Acknowledgments}

I would like to thank Kazunobu Maruyoshi and Kevin Costello for
illuminating discussions, and Petr Va\v{s}ko for careful reading of
the manuscript.  This work is supported by the ERC Starting Grant
No.~335739 ``Quantum fields and knot homologies'' funded by the
European Research Council under the European Union's Seventh Framework
Programme.


\end{document}